\documentclass[longauth]{aa}

\usepackage{hyperref}
\usepackage[usenames,dvipsnames]{xcolor}
\hypersetup{colorlinks=true, 
            citecolor=Cerulean, 
            linkcolor=Cerulean, 
            urlcolor=Cerulean}
\usepackage{makecell}

\usepackage{graphicx}
\usepackage{wrapfig}
\usepackage{subcaption}
\usepackage[rightcaption]{sidecap}
\usepackage{pdflscape}

\usepackage{txfonts}
\usepackage{siunitx}
\usepackage{amsmath}
\usepackage{amssymb}
\usepackage[normalem]{ulem}
\usepackage{lipsum}  
\usepackage{breqn}

\usepackage{natbib}
\usepackage[switch]{lineno} 
\begin{document}

\title{The trans-Neptunian object (119951) 2002 KX14 revealed via multiple stellar occultations}

\author{
J. L. Rizos\inst{1}
\and
J. L. Ortiz\inst{1}
\and
F. L. Rommel\inst{2,3,4}
\and
B. Sicardy\inst{5}
\and    
N. Morales\inst{1}
\and
P. Santos-Sanz\inst{1}
\and
E. Fernández-Valenzuela\inst{2,1}
\and
J. Desmars\inst{6,5}
\and
D. Souami\inst{7,37}
\and
M. Kretlow\inst{1}
\and
A. Alvarez-Candal\inst{1}
\and
J. M. Gómez-Limón\inst{1}
\and
R. Duffard\inst{1}
\and
Y. Kilic\inst{1}
\and
R. Morales\inst{1}
\and
B. J. Holler\inst{8}
\and
M. Vara-Lubiano\inst{1}
\and
A. Marciniak\inst{9}
\and
V. Kashuba\inst{10}
\and
N. Koshkin\inst{10}
\and
S. Kashuba\inst{10}
\and
A. Pal\inst{11}
\and
G. M. Szabó\inst{12}
\and
A. Derekas\inst{12}
\and
L. Szigeti\inst{12}
\and
C. Ellington\inst{13}
\and
O. Schreurs\inst{14}
\and
S. Mottola\inst{15}
\and
R. Iglesias-Marzoa\inst{16}
\and
N. Maícas\inst{16}
\and
F. J. Galindo-Guil\inst{16}
\and
F. Organero\inst{17}
\and
L. Ana\inst{17}
\and
K. Getrost\inst{18}
\and
V. Nikitin\inst{18}
\and
A. Verbiscer\inst{19}
\and
M. Skrutskie\inst{19}
\and
Candace Gray\inst{20}
\and
M. Malacarne\inst{21}
\and
C. Jacques\inst{22}
\and
P. Cacella\inst{23}
\and
O. Canales\inst{24}
\and
D. Lafuente\inst{24}
\and
S. Calavia\inst{24}
\and
Ch. Oncins\inst{24}
\and
M. Assafin\inst{25}
\and
F. Braga-Ribas\inst{3,4}
\and
J. I. B. Camargo\inst{26,4}
\and
A. R. Gomes-Júnior\inst{4,38}
\and
B. Morgado\inst{25}
\and
E. Gradovski\inst{3,4}
\and
R. Vieira-Martins\inst{26,4}
\and
F. Colas\inst{5}
\and
M. Tekes\inst{27,28,29}
\and
O. Erece\inst{30,31}
\and
M. Kaplan\inst{32}
\and
A. Schweizer\inst{33}
\and
J. Kubanek\inst{34,35,36}
}

\institute{
Instituto de Astrofísica de Andalucía – Consejo Superior de Investigaciones Científicas (IAA-CSIC), Glorieta de la Astronomía S/N, E-18008, Granada, Spain
\and
Florida Space Institute, 12354 Research Parkway, Partnership I, Room 211, 32826 Orlando, United States of America
\and
Federal University of Technology - Paraná (UTFPR/PPGFA), Av. Sete de Setembro, 3165, CEP 80230-901 - Curitiba - PR - Brazil
\and
Laboratório Interinstitucional de e-Astronomia (LIneA/INCT do e-Universo), Av. Pastor Martin Luther King Jr, 126, 20765, Rio de Janerio, Brazil
\and
LTE, Observatoire de Paris, Université PSL, Sorbonne Université, Université de Lille, LNE,CNRS 61 Avenue de l’Observatoire, 75014 Paris, France
\and
Institut Polytechnique des Sciences Avancées IPSA, 63 boulevard de Brandebourg, F-94200 Ivry-sur-Seine, France
\and
LIRA, CNRS UMR-8254, Observatoire de Paris, Université PSL, Sorbonne Université, Université Paris Cité, CY Cergy Paris Université, Meudon, 92190, France
\and
Space Telescope Science Institute, 3700 San Martin Drive, Baltimore, USA
\and
Astronomical Observatory Institute, Faculty of Physics and Astronomy, Adam Mickiewicz University, S{\l}oneczna 36, 60-286 Pozna{\'n}, Poland
\and
Astronomical Observatory of Odesa I.I. Mechnikov National University, 1v Marazliivska str., Odesa, 65014, Ukraine
\and
Konkoly Observatory
\and
ELTE Eötvös Loránd University, Gothard Astrophysical Observatory, Szent Imre h u 112, Szombathely, Hungary
\and
Maastricht University, Maastricht Science Programme
\and
Observatoire de Nandrin, Société Astronomique de Liège, Belgium
\and
Institute of Planetary Research, DLR, Rutherfordstr. 2, 12489, Berlin, Germany
\and
Centro de Estudios de Física del Cosmos de Aragón, Plaza de San Juan, 1, 44001, Teruel, Spain
\and
La Hita Observatory
\and
International Occultation Timing Association (IOTA), United States of America
\and
Department of Astronomy, University of Virginia, Charlottesville, VA,  United States of America
\and
New Mexico State University - Apache Point Observatory
\and
GOA, UFES, Av. Fernando FErrari, 514, Vitória, ES, Brazil
\and
Sonear3 Observatory
\and
DogsHeaven Observatory
\and
Botorrita Observatory
\and
Universidade Federal do Rio de Janeiro - Observatório do Valongo, Ladeira Pedro Antônio 43, CEP 20.080-090 Rio de Janeiro, Brazil
\and
Observatório Nacional, Rua Gal. José Cristino 77, Rio de Janeiro RJ-20921-400, Brazil
\and
Space Science and Solar Energy Research and Application Center (UZAYMER), University of Çukurova, Adana 01330, Türkiye
\and
Mesopotamia Astronomy Association, Batman 72040, Türkiye
\and
Yuregir Science Center Adana 01260, Türkiye
\and
National Observatories of Türkiye, TUG, 07058, Antalya, Türkiye
\and
TÜBİTAK, Tunus Cad., No:80, 06680 Kavaklıdere, Ankara, Türkiye
\and
Department of Space Sciences and Technologies, Akdeniz University, Campus, Antalya, 07058,  Türkiye
\and
Buelach Observatory
\and
IOTA/ES, International Occultation Timing Association/European Section e.V., Am Brombeerhag 13, 30459 Hannover, Germany
\and
Czech Astronomical Society, Fričova 298, 251 65, Ondřejov, Czechia
\and
Observatory in Rokycany and Pilsen, p.o., Voldušská 721, 337 01 Rokycany, Czechi
\and
naXys, Department of Mathematics, University of Namur, Rue de Bruxelles 61, Namur 5000, Belgium
\and
Institute of Physics, Federal University of Uberlândia, Av. João Naves de Ávila, Uberlândia, Minas Gerais CEP 38408-100, Brazil
}

\date{Accepted for publication: March 24, 2025}

 \abstract
   {Trans-Neptunian objects (TNOs) are icy bodies located in the outer solar system that offer  key insights into the primordial conditions of our planetary system. The stellar occultation technique has proven to be an essential tool for studying these distant and faint objects, enabling precise determinations of their size, shape, and albedo, while also improving estimates of their orbital parameters. Among them, (119951) 2002 KX14 is a large classical TNO with limited previous observations and unresolved questions regarding its physical properties.}
   {This study aims to analyze and characterize the TNO (119951) 2002 KX14 through observations of stellar occultations, providing unique measurements of its size, shape, and albedo. Shape measurements are especially relevant, as only thirteen TNOs have had their projected shapes measured. These results contribute to our understanding of the physical properties of this object and the broader TNO population.}
   {Five stellar occultations by 2002 KX14 were observed from 2020 to 2023, involving multiple telescopes across different locations in Europe and the Americas. High-precision astrometry and photometric data were used to predict the occultation paths and extract ingress and egress timings. One of the events was detected from six sites and there are also several close misses, providing precise constraints for an accurate determination of the object’s limb. Furthermore, elliptical fits to the occultations chords allowed for the determination of the object's shape and area-equivalent diameter. The geometric albedo was calculated by combining the occultation results with published absolute magnitudes.}
   {The five occultations resulted in 15 positive chords that allowed us to accurately measure the  shape and size of 2002 KX14. Given that the rotational variability of this body is minimal, we can reasonably assume that the variations are due to albedo features, since the body is classed as a Maclaurin spheroid. The projected ellipse has semi-major and semi-minor axes of $241.0 \pm 7.2$ km and $157.1 \pm 5.2$ km, respectively, corresponding to an average area-equivalent diameter of $389.2 \pm 8.7$ km. The geometric albedo was estimated to be $11.9 \pm 0.7\%$. These values differ from the $455 \pm 27$ km diameter and the $9.7^{+1.4}_{-1.3}\%$ albedo derived from thermal measurements, offering a more refined understanding of the object’s physical properties.}
   {}


   \maketitle

\section{Introduction}\label{sec:intro}

Beyond Neptune there lies a vast and diverse population of trans-Neptunian objects (TNOs)\footnote{At the time of writing, more than 5,000 objects, including the scattered disk and Centaurs, have been listed at \href{https://minorplanetcenter.net/iau/lists/TNOs.html}{https://minorplanetcenter.net/iau/lists/TNOs.html} and \href{https://minorplanetcenter.net/iau/lists/Centaurs.html}{https://minorplanetcenter.net/iau/lists/Centaurs.html}.}, which can be dynamically classified into several groups based on their orbital characteristics \citep{Gladman2008}. On the one hand, resonant objects, such as Plutinos, are in a 3:2 resonance with Neptune and follow predictable orbital patterns influenced by the planet’s gravity. On the other hand, the classical belt (nonresonants), is divided into a “cold population,” with low inclinations and nearly circular orbits, and a “hot population,” with higher inclinations and eccentricities. Beyond this lies the scattered disk, containing objects with highly eccentric and inclined orbits that are thought to have been dynamically perturbed by Neptune. Further out, detached objects such as Sedna have perihelia that are far from Neptune’s influence, while some may represent the transition to the hypothesized inner Oort cloud. 

The remote location of these icy bodies, which shields them from the intense solar radiation that affects objects closer to the Sun, makes them valuable objects for studies of the early physical and dynamical conditions that shaped our planetary system. However, it is challenging given their relatively small size (the largest TNO is Pluto, with a radius of 1188 km \citep{Nimmo2017}), along with their low brightness and large heliocentric distances. Moreover, the low surface temperatures of TNOs (30-50 K, e.g. \cite{Stern2008}) cause their thermal emission to peak in the far-infrared (FIR) region of the electromagnetic spectrum, which is blocked by Earth’s atmosphere. This makes space-based observations essential for their study. However, the stellar occultation technique enables ground-based observations and has proven to be one of the most effective methods for studying TNOs to date. This technique has significantly advanced our understanding of these distant icy objects by allowing for highly accurate size measurements, revealing the presence of atmospheres, moons, or rings. In addition, when combined with independent datasets, it helps determine their geometric albedo (e.g. \cite{Hubbard1988,Sicardy2011,Ortiz2017,Assafin2023}). Moreover, it achieves kilometric precision, making it especially useful for refining the orbital parameters through astrometry and making highly accurate shape determinations.

In this work, we focus on the trans-Neptunian object (119951) 2002 KX14, discovered in 2002, which has a semi-major axis of 38.63 au, an eccentricity of 0.04, and an inclination of 0.41°. This object exhibits orbital characteristics; specifically, low eccentricity and low inclination, which are typical of the cold classical population. Members of this population are the most dynamically stable objects in the classical belt and are believed to have formed in situ, making them valuable tracers of the early formation history of the trans-Neptunian region \citep{Delsanti2006}. However, the semi-major axis of 2002 KX14 places it outside the classical belt region, which is generally defined between $\sim$42 and 47 au. Some authors (e.g., \citealt{Estela2021}) have classified it as an inner classical object, a population that is not considered to be an extension of the cold classicals, but instead belonging to the low-inclination tail of the hot population \citep{Vilenius2012}. Unlike the cold classicals, the hot population is thought to have been dynamically excited and relocated to its current position during the migration of the giant planets. During this migration, gravitational interactions perturbed their orbits, increasing both their inclination and eccentricity. Therefore, studying 2002 KX14 is particularly valuable because its orbital properties place it near the boundary between the cold and hot classical populations, offering insights into the dynamical structure and evolutionary history of the trans-Neptunian region.

Based on thermal measurements with the \textit{Herschel} Space Telescope, \cite{Vilenius2012} estimated a diameter of $455 \pm 27$ km and a geometric albedo ($p_{V})$ of $9.7\substack{+1.4 \\ -1.3 } \%$. This diameter is close to the size where hydrostatic equilibrium may no longer be maintained \citep{Tancredi2008}. Subsequent ground-based observations studied its photometric properties, providing values of its absolute magnitude in the R-band: $H_{R} = 4.349 \pm 0.124$ \citep{Peixinho2012}, $H_{R} = 4.46 \pm 0.01$ \citep{Benecchi2013}, and in the V-band: $H_{V} = 5.07 \pm 0.03$ \citep{Boehnhardt2014}, $H_{V} = 4.978 \pm 0.017$ \citep{Alvaro2016}. Also, \cite{Alvaro2014} recorded a positive occultation, measuring a chord of $415 \pm 1$ km and retrieving an equivalent diameter of $365\substack{+30 \\ -21 }$ km. Subsequently, \cite{Rommel2020} obtained two additional chords of $421 \pm 3$ km and $431 \pm 6$ km.  Despite these successful observations, due to its faint apparent magnitude above 20.5 in the V-band, with rotational variations under 0.05 mag \citep{Benecchi2013}, the shape of 2002 KX14 as well as its pole and rotation period are unknown. 

In this study, we present the results of five positive occultations, which yield a total of 15 chords on five different dates. These observations have allowed us to determine the object's shape and geometric albedo. Currently, only thirteen TNOs have projected shape determinations \citep{Ortiz2012,Benedetti-Rossi2016,Ortiz2017,Dias-Oliveira2017,Benedetti-Rossi2019,Ortiz2020,Souami2020,Santos-Sanz2022,Vara-Lubiano2022,Pereira2023, Rommel2023, Kretlow2024}, with 2002 KX14 being now the first\footnote{Apart from Arrokoth, which was observed up close by the New Horizons mission} cold classical object (if confirmed). The fact that it is close to the critical diameter beyond which hydrostatic equilibrium may no longer be maintained makes determining its shape a particularly interesting undertaking. Therefore, this study is poised to contribute to the growing body of knowledge on TNOs, laying the groundwork for future in-depth studies, providing data to enhance the statistics that allow us to contextualize this population more comprehensively, and demonstrating the effectiveness of the occultation technique in characterizing these distant objects.

\section{Observations}\label{sec:observations}

\begin{table*}[t]
\centering
\caption{Occulted stars by 2002 KX14 presented in this work.}
\label{table:stars}
\renewcommand{\arraystretch}{1.5}
\begin{tabular}{c c c c c}

        \textbf{Occultation}      & \textbf{Source (Gaia-DR3)}      &  \textbf{Right ascension (ICRF)}  & \textbf{Declination (ICRF)} & \textbf{G, V, K magnitudes} \\
        \hline
        
        A & 4111560308371475840  & 17 24 20.6730 & -23 21 50.203 & 14.54, 14.75, 10.52 \\

        B & 4116601049409385472  & 17 35 07.9754 & -23 28 26.010 & 16.82, 17.78, 12.31 \\

        C & 4068746489444765440   & 17 45 03.2702 & -23 32 34.722 & 15.55, 16.47, 10.71 \\

        D & 4068793394867887872   & 17 44 54.9895  & -23 32 31.595 & 16.28, 16.39, 11.36 \\

        E & 4116448256084153472   & 17 41 01.9529  & -23 30 54.310 & 16.15, 15.90, 11.90 \\
       
\end{tabular}
\tablefoot{Stellar right ascension (h m s) and declination (d m s) represent the astrometric coordinates on the International Celestial Reference Frame (ICRF) of the stars, corrected of proper motion, given for the moments of the occultations.}

\end{table*}

Initial predictions for 2002 KX14 were made using the Numerical Integration of the Motion of an Asteroid (NIMA) solution \citep{Desmars2015}. Subsequently, several observational campaigns were conducted prior to each occultation using the 1.23 m telescope at the Calar Alto Observatory and the 2.0 m Liverpool Telescope located at the Roque de los Muchachos Observatory. These efforts were aimed at performing astrometry for this TNO and refining the predicted occultation shadow path, which was communicated to observers for proper deployment, in cases where mobile equipment was used. This approach enabled the observation of five positive occultations in 2020, 2022, and 2023, as described in the following sections.

\subsection{Occultation A: May 26, 2020}\label{sec:stars_occA}

On May 26, 2020, an occultation by 2002 KX14 of the star Gaia DR3 4111560308371475840 (see Table \ref{table:stars}) was predicted to happen at 22:29:12 UT (referred to hereafter as "Occ. A").

\begin{figure*}[h!]
\centering

\begin{subfigure}[t]{0.43\linewidth}
\centering
\captionsetup{justification=raggedright,singlelinecheck=false}
\caption{}
\includegraphics[width=\linewidth]{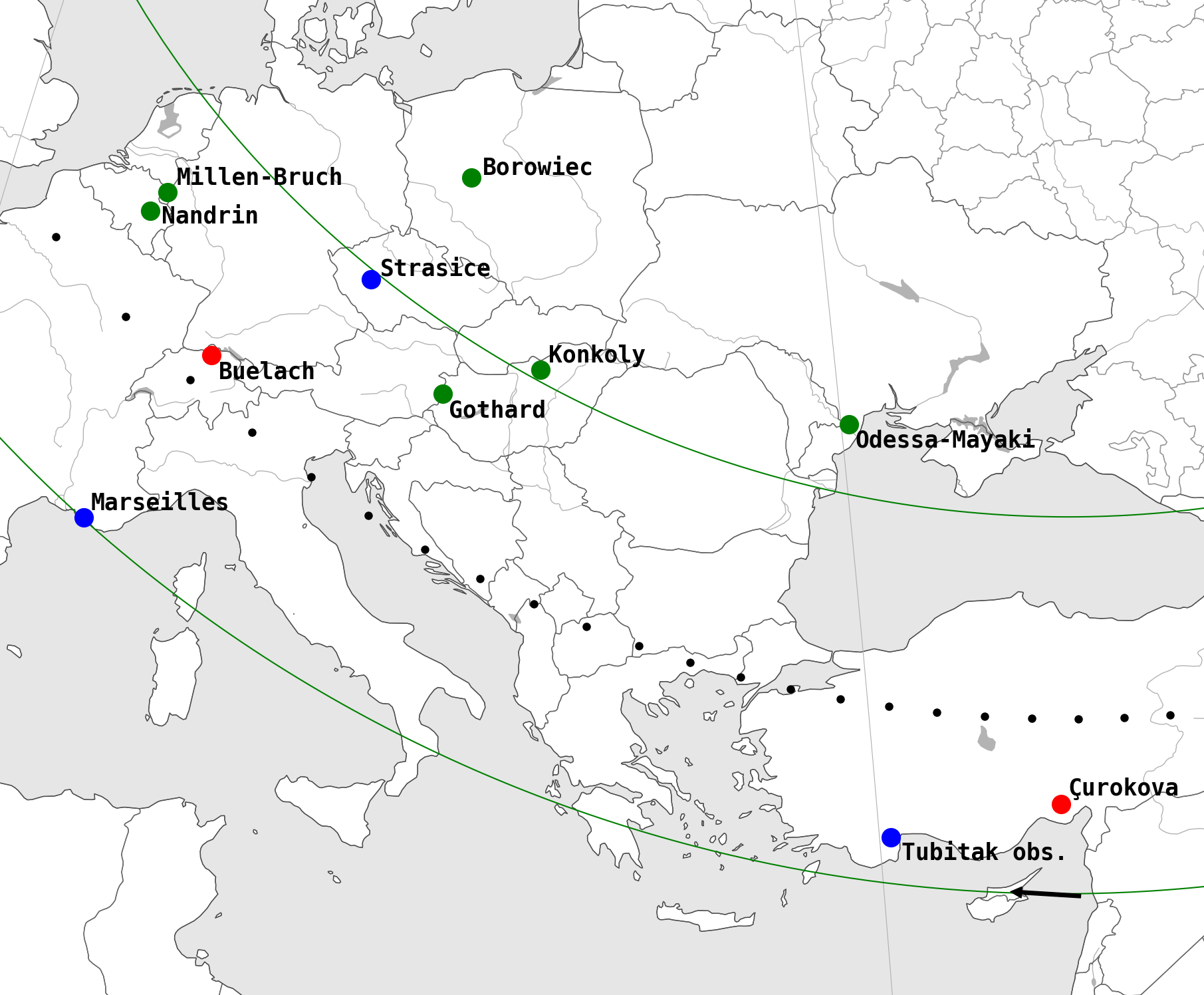}\label{fig:shadowsA} 
\end{subfigure}
\hfill
\begin{subfigure}[t]{0.43\linewidth}
\centering
\captionsetup{justification=raggedright,singlelinecheck=false}
\caption{}
\includegraphics[width=\linewidth]{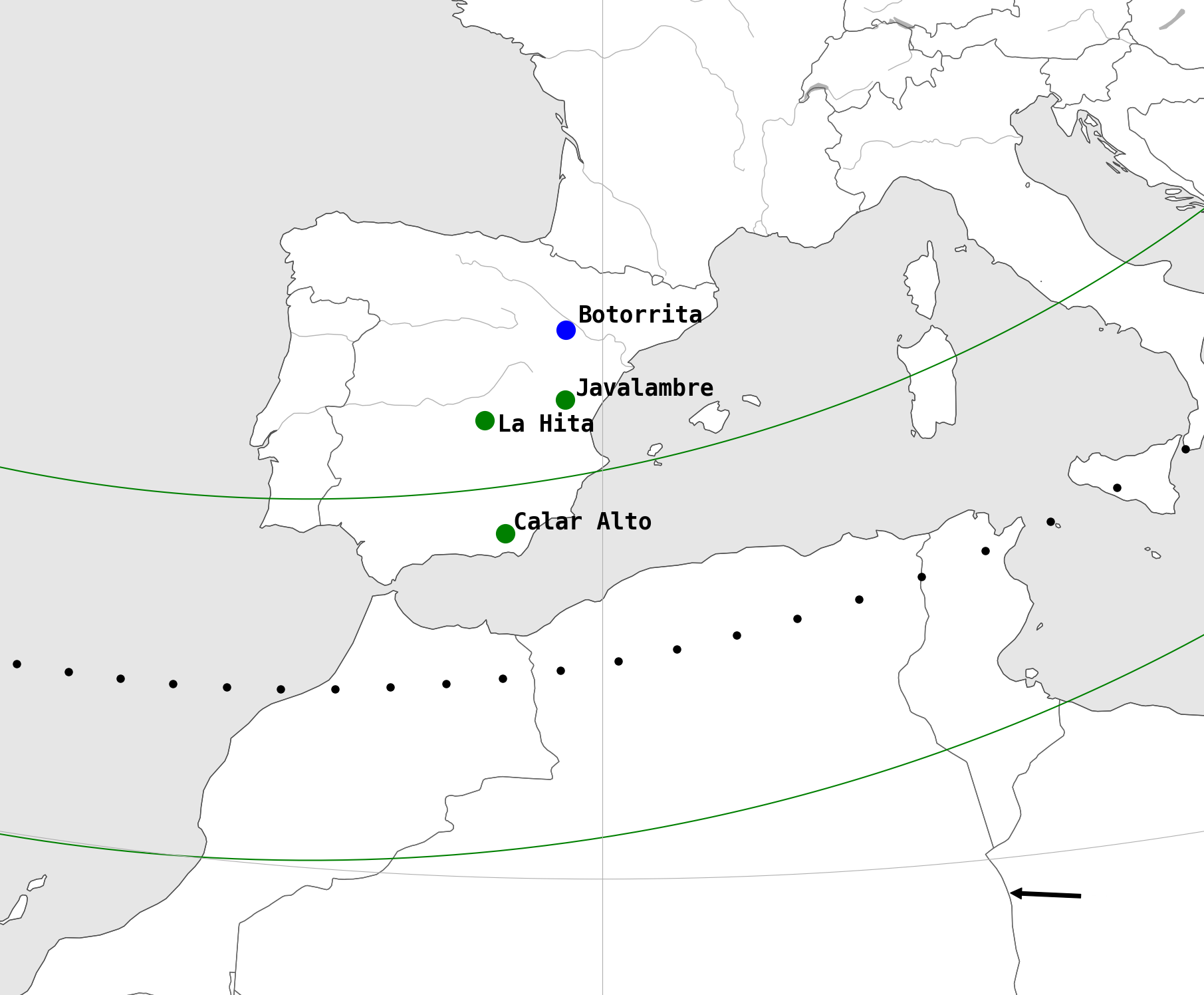}\label{fig:shadowsB} 
\end{subfigure}

\medskip  

\begin{subfigure}[t]{0.43\linewidth}
\centering
\captionsetup{justification=raggedright,singlelinecheck=false}
\caption{ }
\includegraphics[width=\linewidth]{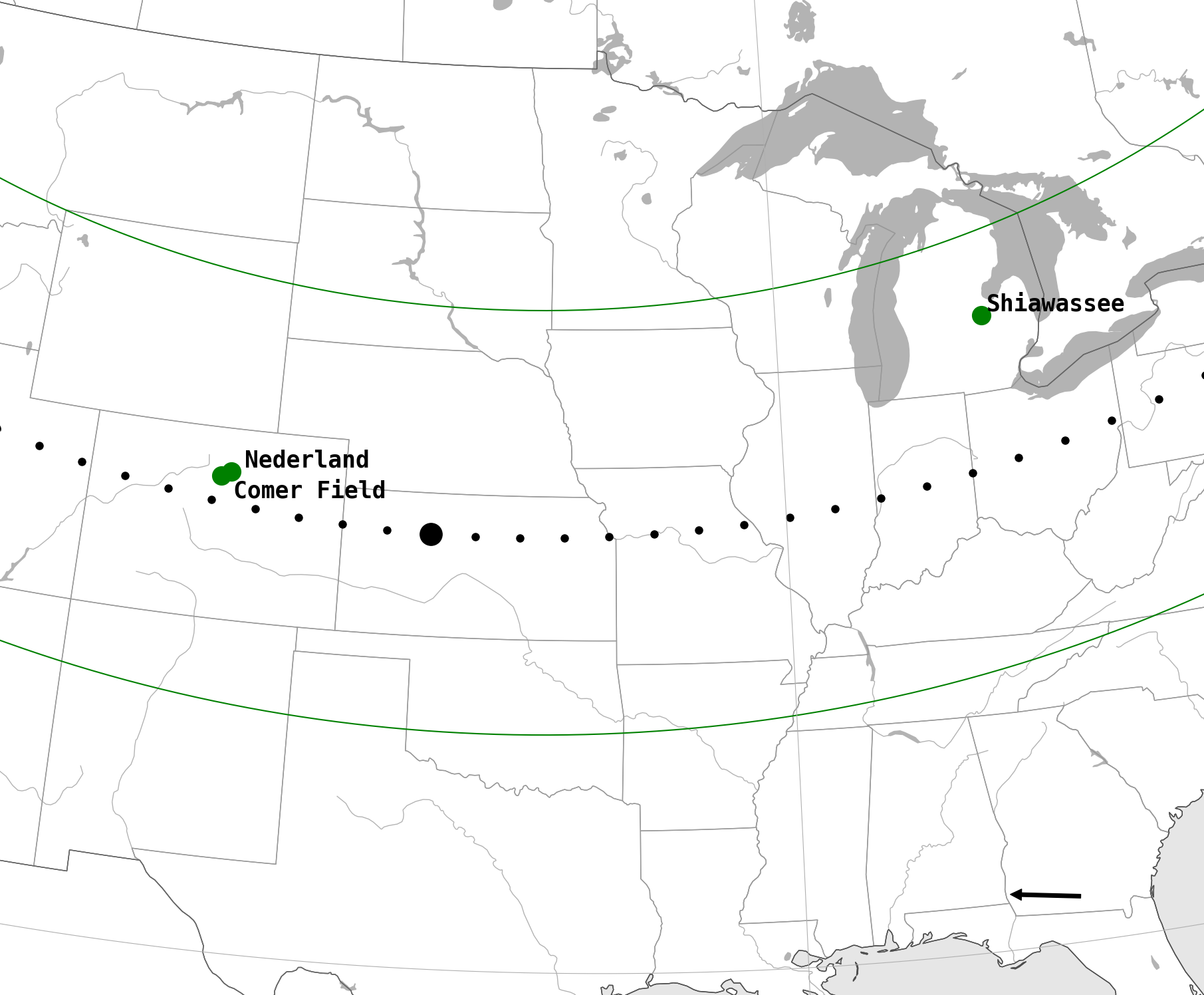}\label{fig:shadowsC} 
\end{subfigure}
\hfill
\begin{subfigure}[t]{0.43\linewidth}
\centering
\captionsetup{justification=raggedright,singlelinecheck=false}
\caption{ }
\includegraphics[width=\linewidth]{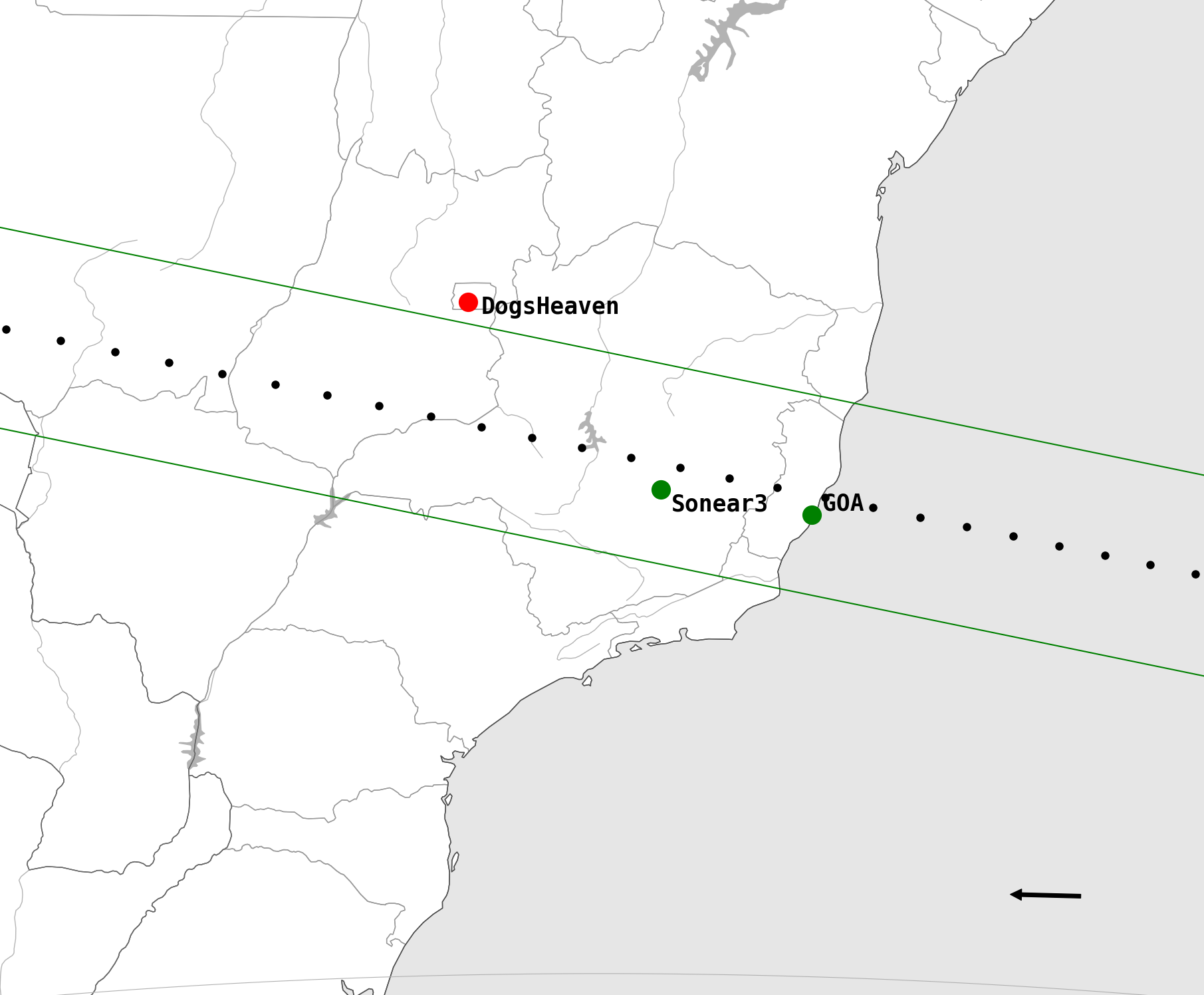}\label{fig:shadowsD} 
\end{subfigure}

\medskip  

\begin{subfigure}[t]{0.43\linewidth}
\centering
\captionsetup{justification=raggedright,singlelinecheck=false}
\caption{ }
\includegraphics[width=\linewidth]{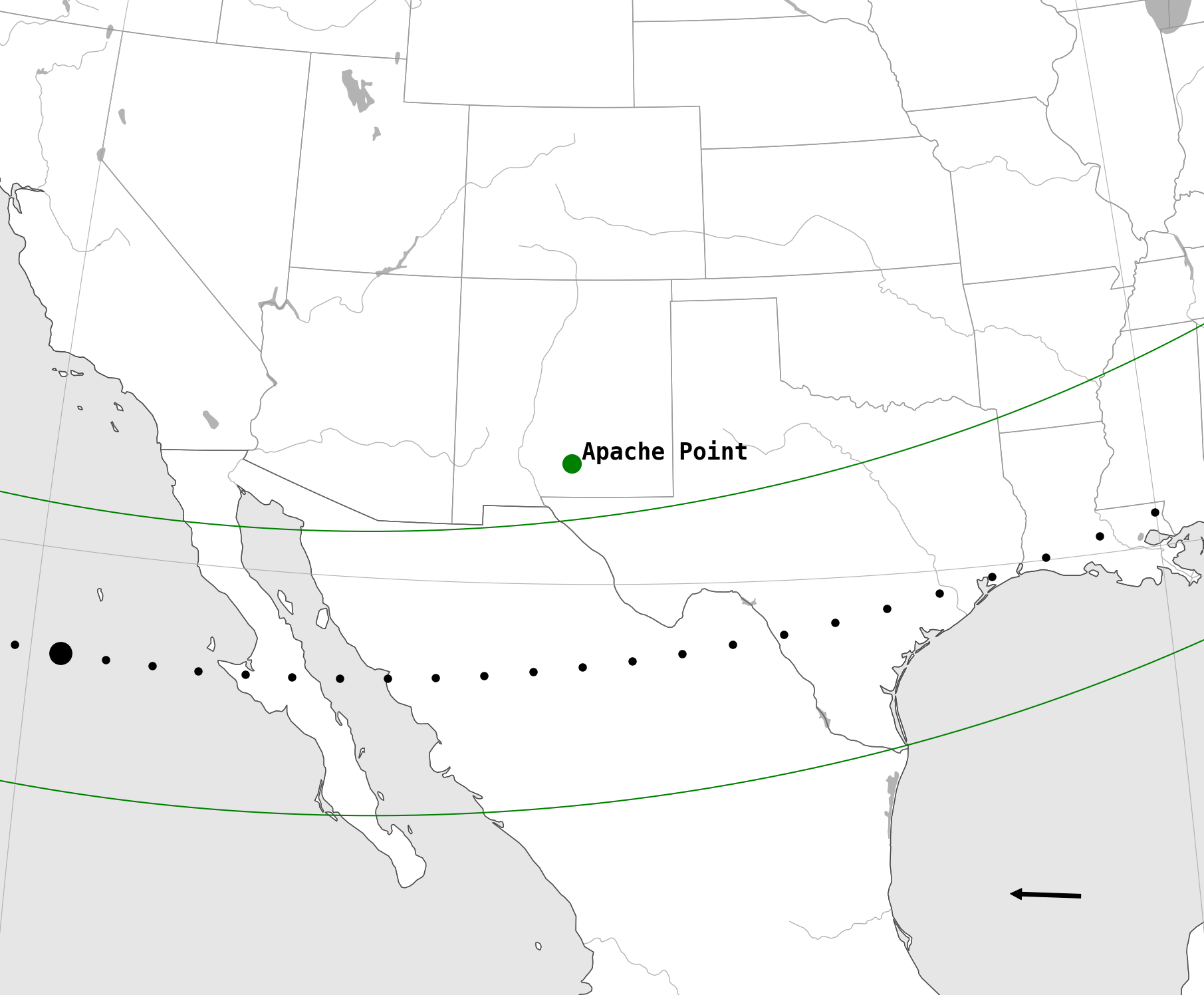}\label{fig:shadowsE} 
\end{subfigure}

\caption{\label{fig:shadows} 
Predicted occultation shadow path computed from JPL$\#$18 ephemerids for (a) May 26, 2020, (b) June 23, 2022, (c)  May 17, 2023, (d) May 19, 2023, and (e) July 01, 2023. The green lines depict the boundaries of the shadow path assuming a 455 km diameter body. The black points indicate the center of the shadow path at 5-second intervals, with the largest point representing the central moment of the prediction. The dots indicate the positions of the stations involved in the campaigns. Green dots signify sites where a positive occultation was observed, red dots a negative occultation, and blue ones indicate low S/N or technical issues (neither positive nor negative). These maps have been generated using SORA. The bottom right arrows denote the shadow's direction of motion.}
\end{figure*}

The event was observed from nine different locations, resulting in six positive detections from Borowiec (Poland), Odesa-Mayaki Observatory (Ukraine), Konkoly Observatory (Hungary), Gothard Astrophysical Observatory (Hungary), Millen-Bruch (Germany), and Nandrin (Belgium); and two negative detections: from Buelach (Switzerland) and Çukurova (Turkey). In addition, it was also observed from Tubitak Observatory (Turkey) and Marseilles (France), but the signal-to-noise ratio (S/N) of the data was too low to confirm whether the occultation occurred or not. An additional positive detection was obtained from Strasice (Czechia), but due to technical issues with the data format, it was not possible to determine the ingress and egress times with sufficient accuracy to be useful for this work. Fig. \ref{fig:shadowsA} shows the predicted occultation shadow path along with the position of the observing stations involved in this work. Table \ref{table:observatories} expands on the information regarding locations, instrumentation, and the observers involved. Time series observations were obtained and time synchronized using Network Time Protocol (NTP) servers or global positioning systems (GPS) devices. Image acquisition started at least $\sim$5 min before and ended $\sim$5 min after the occultation time. No filters were used to maximize the S/N of the occulted star.

The angular diameter of the occulted star was estimated using the following equation \citep{vanBelle1999}:

\begin{equation}\label{eq:vanbelle}
    \theta =\frac{ 10^{a+b(V-K)}}{10^{V/5}},
\end{equation}

\begin{figure*}[h!]
        \centering
        \includegraphics[width=0.5\linewidth]{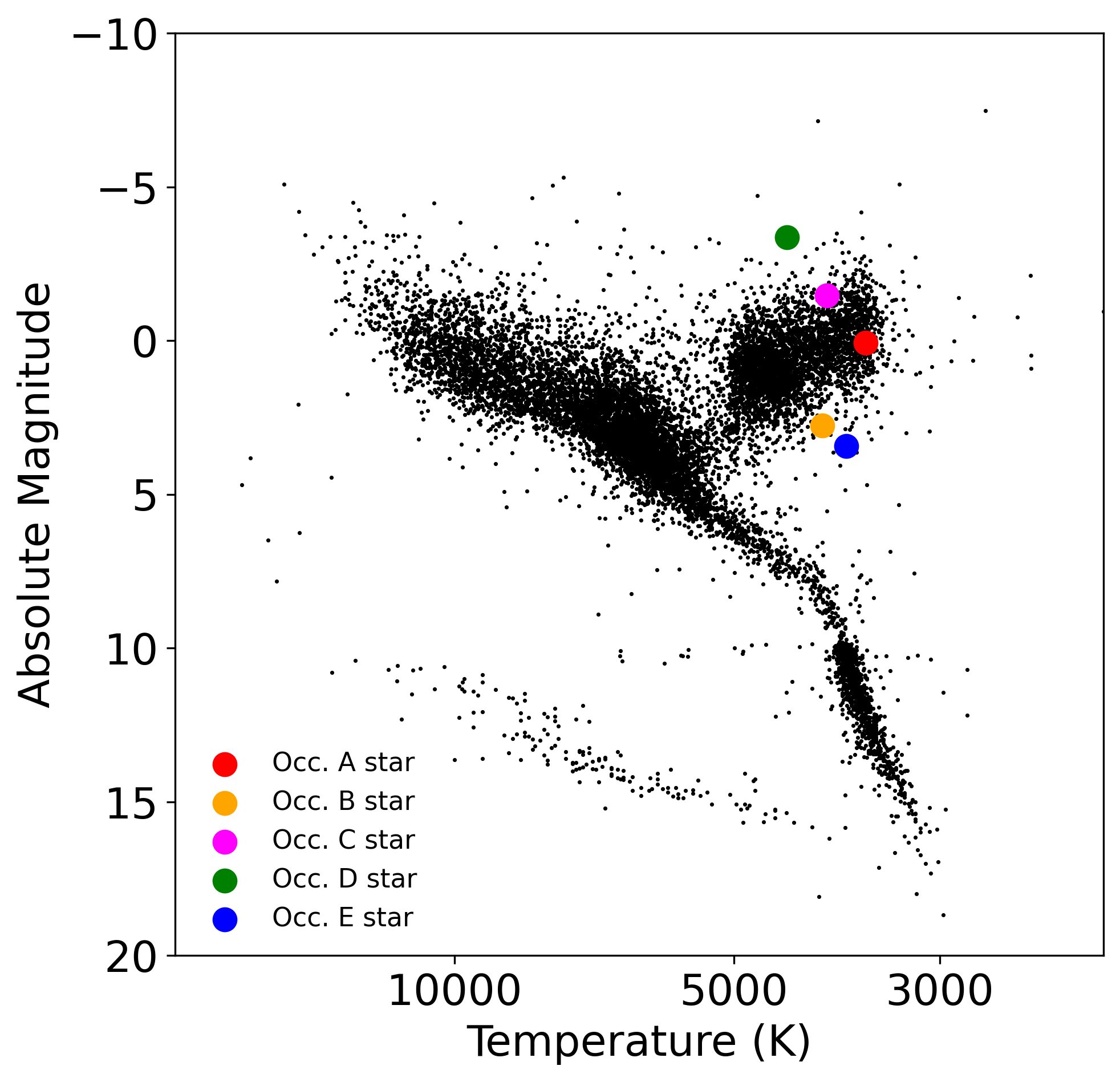}
        \caption{Absolute magnitude versus temperature of the HYG star database along with the stars occulted by 2002 KX14 and analyzed in this work.}\label{fig:HR_diagram}
\end{figure*}

As it is a giant star (see the Hertzsprung–Russell diagram in Fig. \ref{fig:HR_diagram})  $(a, b) = (0.669, 0.223)$ according to \cite{vanBelle1999}. $V$ and $K$ denote the apparent magnitudes of the star in the V-band and K-band, being $V$ = 14.750 mag and $K$ = 10.520 mag \citep{Zacharias2004}. Consequently, the angular diameter is calculated as 0.046 mas, corresponding to 1.26 km at a 2002 KX14's geocentric distance ($\Delta$) of 38.01 au. As the central wavelength ($\lambda$) is $\sim$600 nm for our CCD/CMOS observations, using Eq. \ref{eq:fresnel}, the Fresnel scale of this object can be calculated as 1.3 km,

\begin{equation}\label{eq:fresnel}
    F =\sqrt{\frac{\lambda \Delta}{2}}.
\end{equation}

The minimum cycle time (exposure plus read-out time) here comes from the observations made in Gothard Observatory and Millen-Bruch, which was recorded at 0.7 seconds. With 2002 KX14's velocity at the time of occultation measured at 23.41 km/s, this cycle time corresponds to a distance of 16.38 km. Given that this distance is an order of magnitude larger than that derived from the star's angular size and Fresnel diffraction effects, they have a small impact on the derivation of ingress and egress times. Nevertheless, they were taken into account.

\subsection{Occultation B:\ June 23, 2022}

On June 23, 2022, an occultation of the star Gaia DR3 4116601049409385472 by 2002 KX14 (Table \ref{table:stars}), with a predicted maximum duration of 18.7 seconds, was forecasted to occur at 00:51:33 UT (hereafter, "Occ. B"). The event was observed from four locations in Spain: the Botorrita, Javalambre, La Hita, and Calar Alto observatories. Three of these observatories, Javalambre, La Hita, and Calar Alto, reported positive detections, while Botorrita experienced technical issues (Fig. \ref{fig:shadowsB}, Table \ref{table:observatories}). Time series observations were captured and synchronized using either NTP servers or GPS devices. Image acquisition started approximately $\sim$5 minutes before and concluded $\sim$5 minutes after the predicted event time. No filters were applied to optimize the S/N of the star.

The angular diameter of the occulted star was estimated to be 0.019 mas using Eq. \ref{eq:vanbelle} (giant star, see Fig. \ref{fig:HR_diagram}), with magnitudes of $V = 17.78 $ and $K = 12.31 \text{ mag}$ from \cite{Zacharias2004,Lasker2008}. At 2002 KX14's geocentric distance of $\Delta = 37.88$ au, this corresponds to a physical diameter of 0.58 km. The Fresnel scale in this case is 1.3 km. Given that the minimum observation cycle time was 1.5 seconds and 2002 KX14’s velocity during the occultation was 24.33 km/s, this timing corresponds to a distance of 36.49 km. Thus, the ingress and egress determination uncertainty mainly comes the photometric accuracy of the occultation light curves. Still, the star's angular size and Fresnel diffraction effects were included in the analysis.

\subsection{Occultation C:\ May 17, 2023}

On May 17, 2023, an occultation by 2002 KX14 of the star Gaia DR3 4068746489444765440 (Table \ref{table:stars}) with a maximum duration of 22.3 seconds was predicted at 08:51:46 UT (referred to hereafter as "Occ. C"). The star was observed from three telescopes, resulting in all cases with positive detections. The telescopes were located in Shiawassee, Comer Field and Nederland, USA (Fig. \ref{fig:shadowsC}, Table \ref{table:observatories}). Time series observations were obtained and synchronized using GPS devices. Image acquisition started at least $\sim$5 min before and ended $\sim$5 min later than the predicted time. No filters were used to maximize the S/N of the star.

An angular diameter of 0.046 mas of the occulted star was estimated using Eq. \ref{eq:vanbelle} (giant star, see Fig. \ref{fig:HR_diagram}), with $V$ = 16.47 mag and $K$ = 10.71 mag from \cite{Zacharias2004}. This corresponds to 1.25 km at 2002 KX14's geocentric distance, $\Delta$, of 37.97 au. The Fresnel scale in this case is 1.3 km. Because the minimum cycle time for these observations is 2.0 seconds, while 2002 KX14's velocity at the time of occultation was  20.38 km/s, it translates to a distance of 40.76 km. Given that this distance is orders of magnitude greater than that derived from the star's angular size and Fresnel diffraction effects, the stellar angular size and diffraction have a slight impact on the derivation of ingress and egress times. However, they were accounted for.

\subsection{Occultation D:\ May 19, 2023}

Two days later, on May 19, 2023, an occultation by 2002 KX14 of the star Gaia DR3 4068793394867887872 (Table \ref{table:stars}) with a maximum duration of 21.8 seconds was predicted to happen at 03:08:49 UT (hereafter "Occ. D"). Observations of this event were conducted from three locations in Brazil, with positive detections reported by the GOA and Sonear3 observatories; whereas in DogsHeaven, the S/N of the data was too low (Fig. \ref{fig:shadowsD}, Table \ref{table:observatories}). Time series data were collected and synchronized using NTP servers. The image capture began approximately $\sim$5 min before the event and continued for about five minutes afterward. To maximize the star's S/N, no filters were applied.

Using Eq. \ref{eq:vanbelle} (giant star; see Fig. \ref{fig:HR_diagram}), the occulted star’s diameter was estimated to be 0.032 mas, with magnitudes of $V$ = 16.39 mag and $K$ = 11.36 mag from \cite{Zacharias2004}. At 2002 KX14's geocentric distance of 37.96 au, this angular size corresponds to a physical diameter of 0.89 km. The Fresnel scale is again 1.3 km. Given that the minimum observation cycle time is 6.8 seconds and that 2002 KX14's velocity during the occultation was 20.83 km/s, this timing translates to a spatial distance of 141.64 km. Thus, the ingress and egress determination uncertainty mainly comes from the photometric accuracy of the occultation light curves. Even so, the star's angular size and Fresnel diffraction effects were included in the analysis.

\subsection{Occultation E:\ July 01, 2023}

On July 01, 2023, the last occultation occultation by 2002 KX14 of the star Gaia DR3 4116448256084153472 (Table \ref{table:stars}), with a maximum duration of 19.2 seconds, was predicted to occur at 07:09:22 UT (referred to hereafter as “Occ. E”). The star was observed from a single telescope, which resulted in a positive detection. The telescope was located in the Sacramento Mountains, New Mexico, USA (Fig. \ref{fig:shadowsE}, Table \ref{table:observatories}). Time-series observations were obtained and synchronized using GPS devices. Image acquisition started at least $\sim$5 minutes before and ended $\sim$5 minutes after the predicted time. No filters were used to maximize the S/N of the star.

An angular diameter of 0.04 mas for the occulted star was estimated using Eq. \ref{eq:vanbelle} (giant star, see Fig. \ref{fig:HR_diagram}), with $V$ = 16.15 mag and $K$ = 11.90 mag from \cite{Zacharias2004}. This corresponds to 0.65 km at 2002 KX14’s geocentric distance, $\Delta$, of 37.85 au. The Fresnel scale in this case is again 1.3 km. Since the cycle time for this observation was 0.5 seconds, while 2002 KX14’s velocity at the time of occultation was 23.74 km/s, this translates to a distance of 11.87 km. Given that this distance is several orders of magnitude greater than that derived from the star’s angular size and Fresnel diffraction effects, the stellar angular size and diffraction have a negligible impact on the derivation of ingress and egress times. Even so, they still were taken into consideration.

\section{Data analysis and results}\label{sec:data_analysis}

\subsection{Stellar occultations}

The data were compiled and managed via the Tubitak Occultation Portal website \citep{Kilic2022}.  Initially, CCD/CMOS image sequences were calibrated for bias and flat-field following standard procedures with the AstroImageJ (AIJ) software \citep{Collins2017} when they were available. For observations recorded in video format, we used the Planetary Imaging PreProcessor (PIPP\footnote{\url{https://pipp.software.informer.com}}) software to convert the video files to FITS format before proceeding with photometric analysis.

\begin{figure*}[h!]
\section{Stellar occultation analyses}

Light curves from the occultation events described in Section \ref{sec:data_analysis}. Additionally, tables containing the ingress and egress times after the square well fits are provided, along with the expanded table containing information about the location of telescopes, instruments, and observers.

\centering

\begin{subfigure}[t]{0.33\linewidth}
\centering
\captionsetup{justification=raggedright,singlelinecheck=false}
\caption{}
\includegraphics[width=\linewidth]{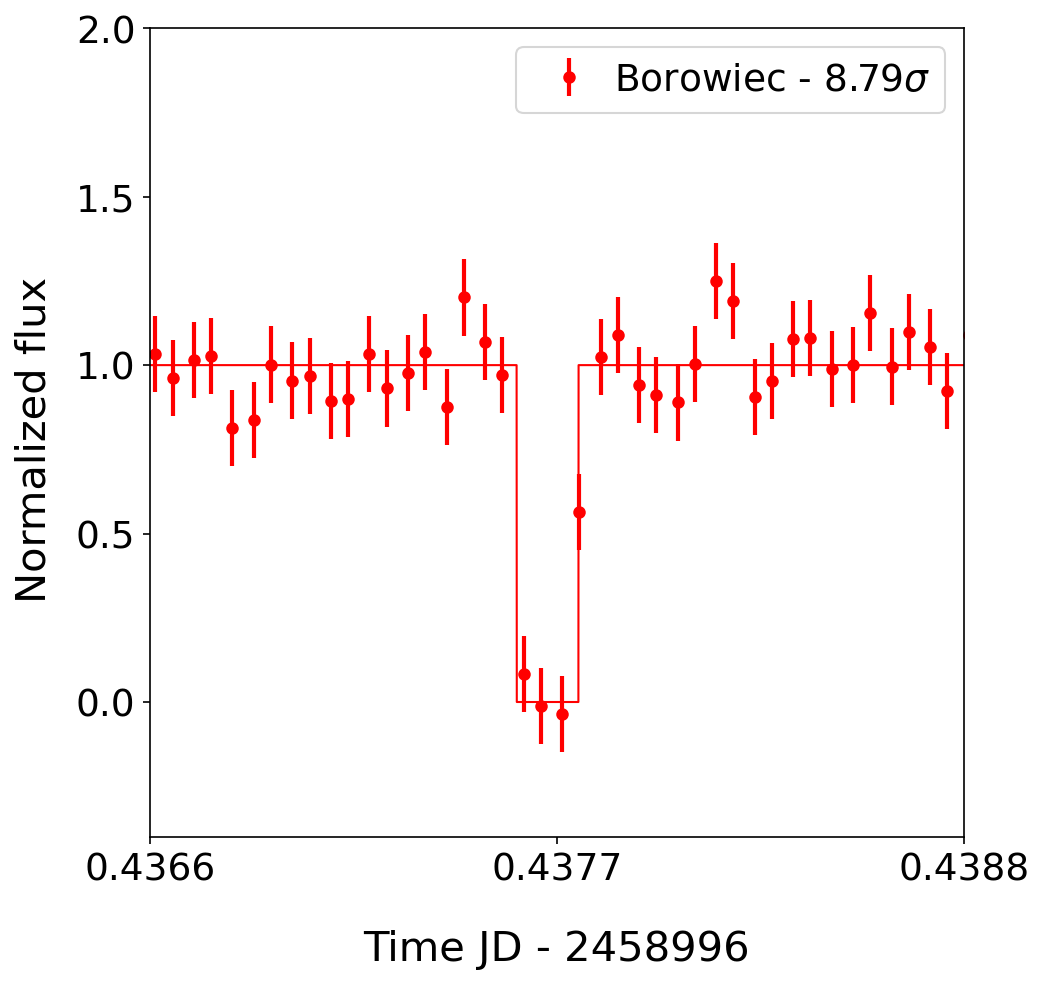} 
\end{subfigure}
\hfill
\begin{subfigure}[t]{0.33\linewidth}
\centering
\captionsetup{justification=raggedright,singlelinecheck=false}
\caption{}
\includegraphics[width=\linewidth]{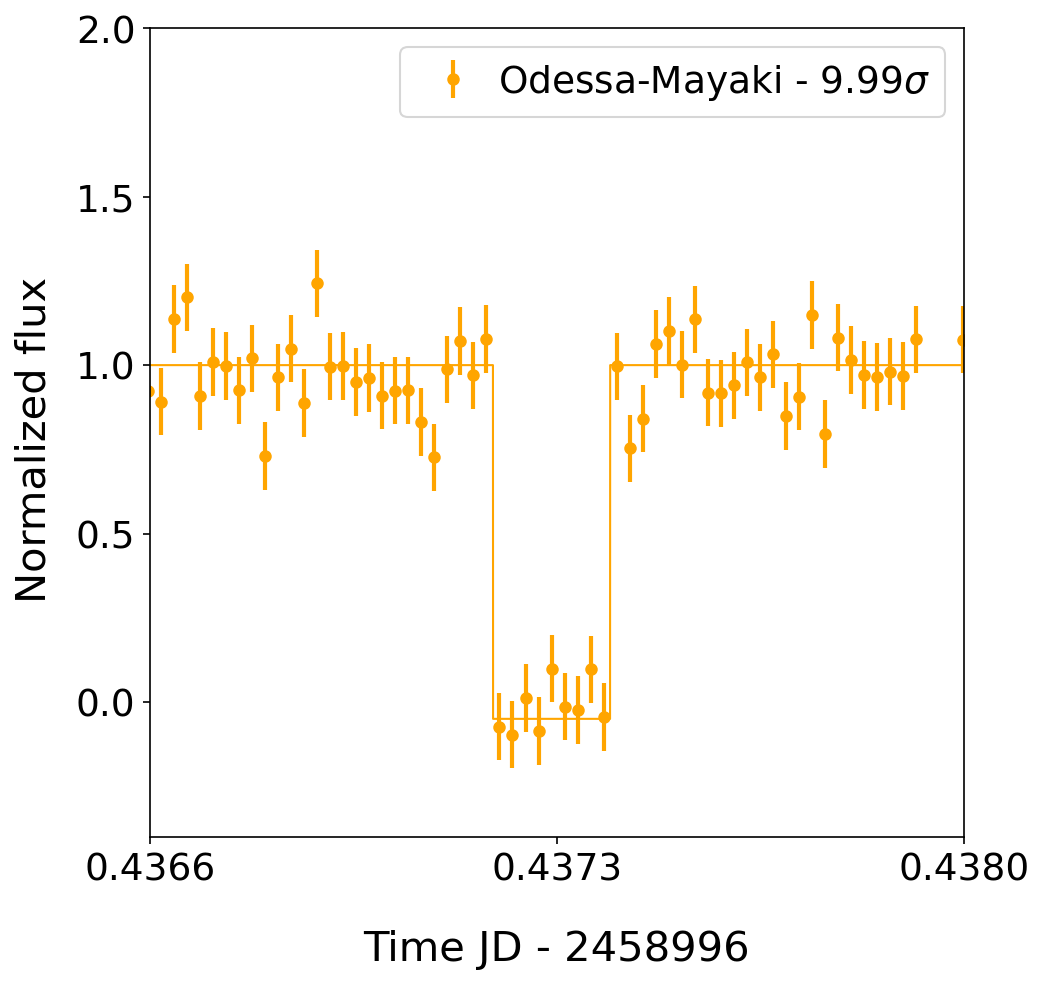}
\end{subfigure}
\medskip  
\begin{subfigure}[t]{0.33\linewidth}
\centering
\captionsetup{justification=raggedright,singlelinecheck=false}
\caption{ }
\includegraphics[width=\linewidth]{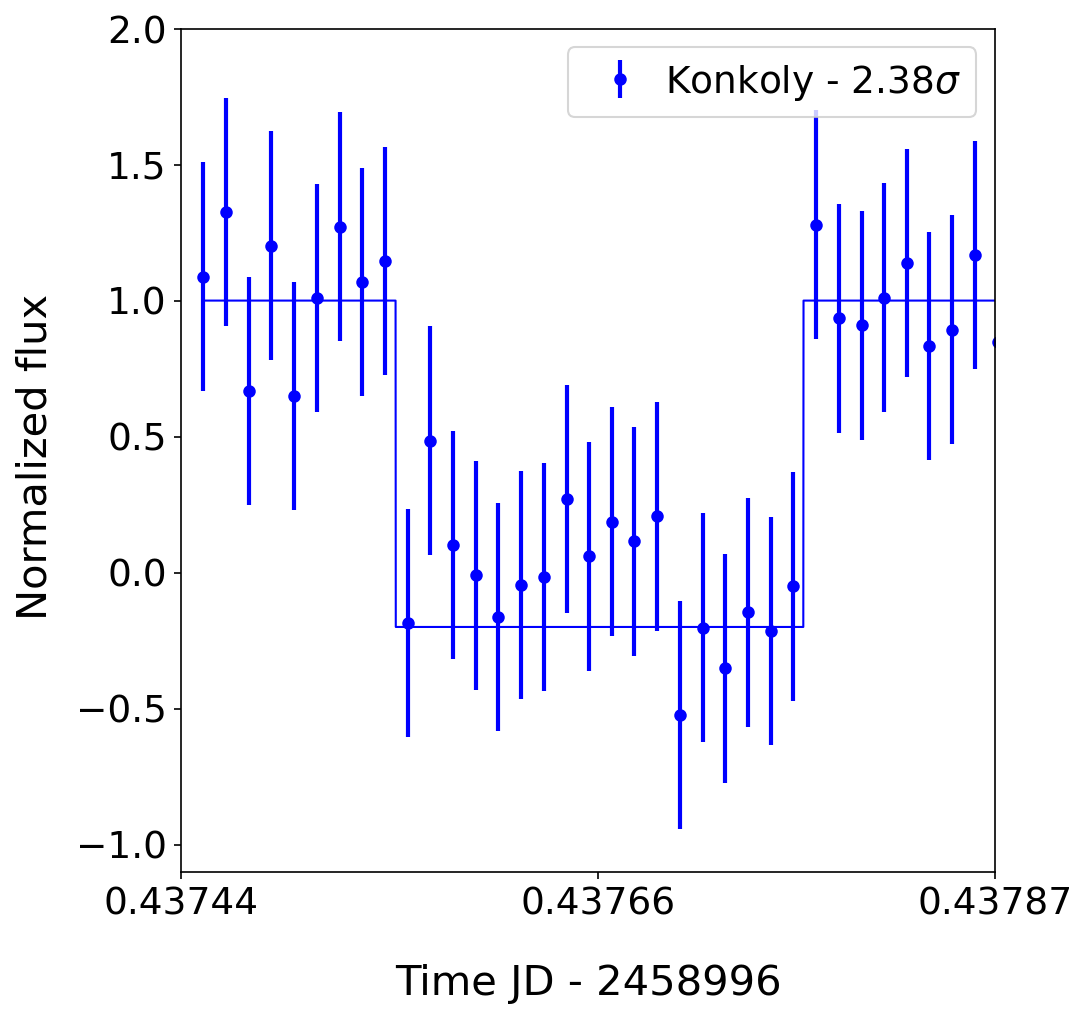}
\end{subfigure}
\hfill
\begin{subfigure}[t]{0.33\linewidth}
\centering
\captionsetup{justification=raggedright,singlelinecheck=false}
\caption{ }
\includegraphics[width=\linewidth]{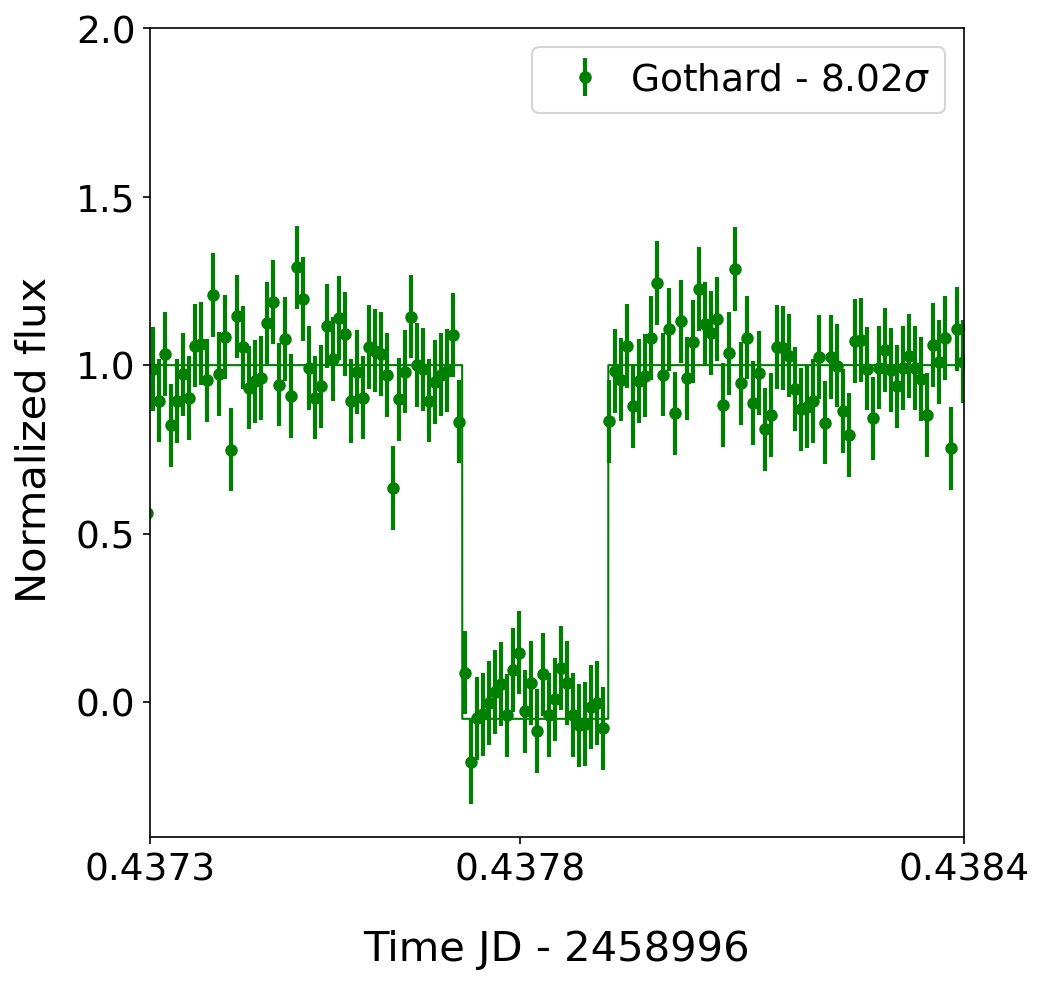}
\end{subfigure}
\medskip  
\begin{subfigure}[t]{0.33\linewidth}
\centering
\captionsetup{justification=raggedright,singlelinecheck=false}
\caption{ }
\includegraphics[width=\linewidth]{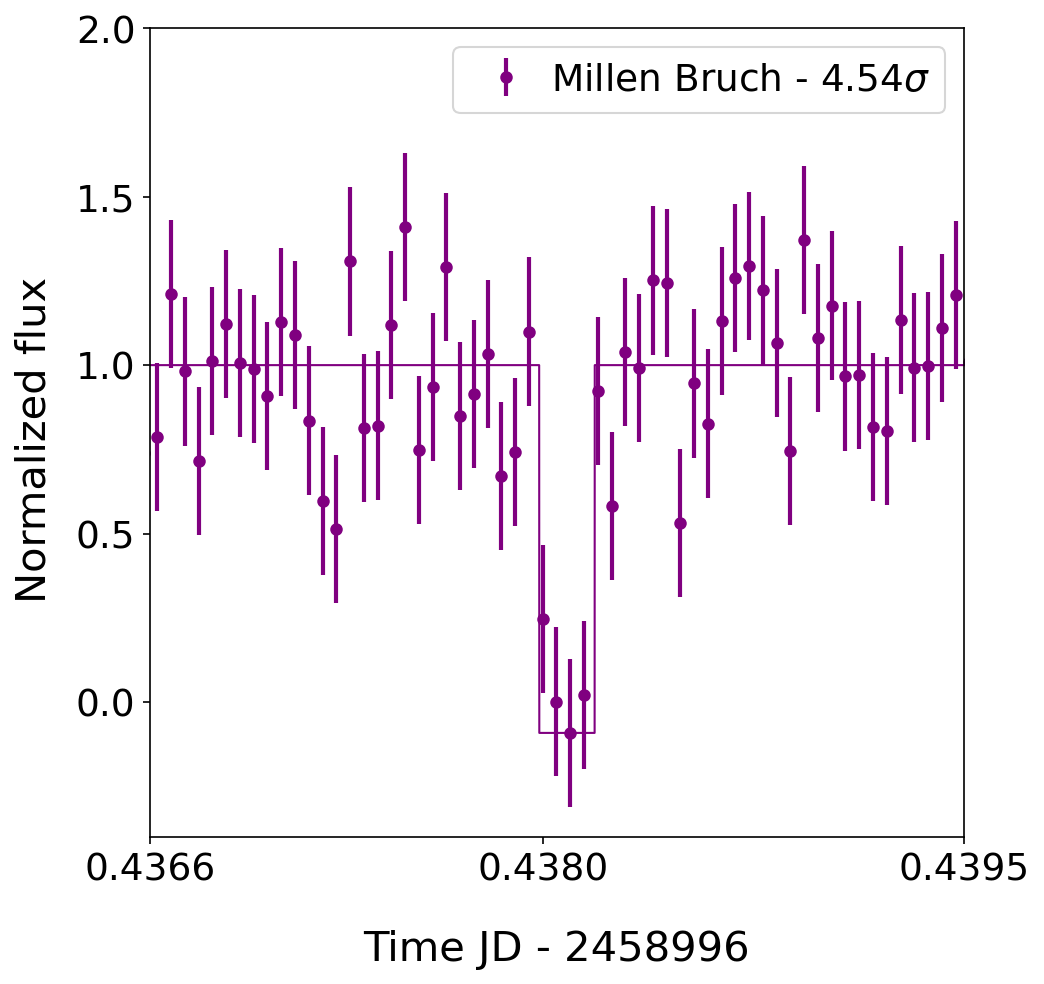}
\end{subfigure}
\hfill
\begin{subfigure}[t]{0.33\linewidth}
\centering
\captionsetup{justification=raggedright,singlelinecheck=false}
\caption{ }
\includegraphics[width=\linewidth]{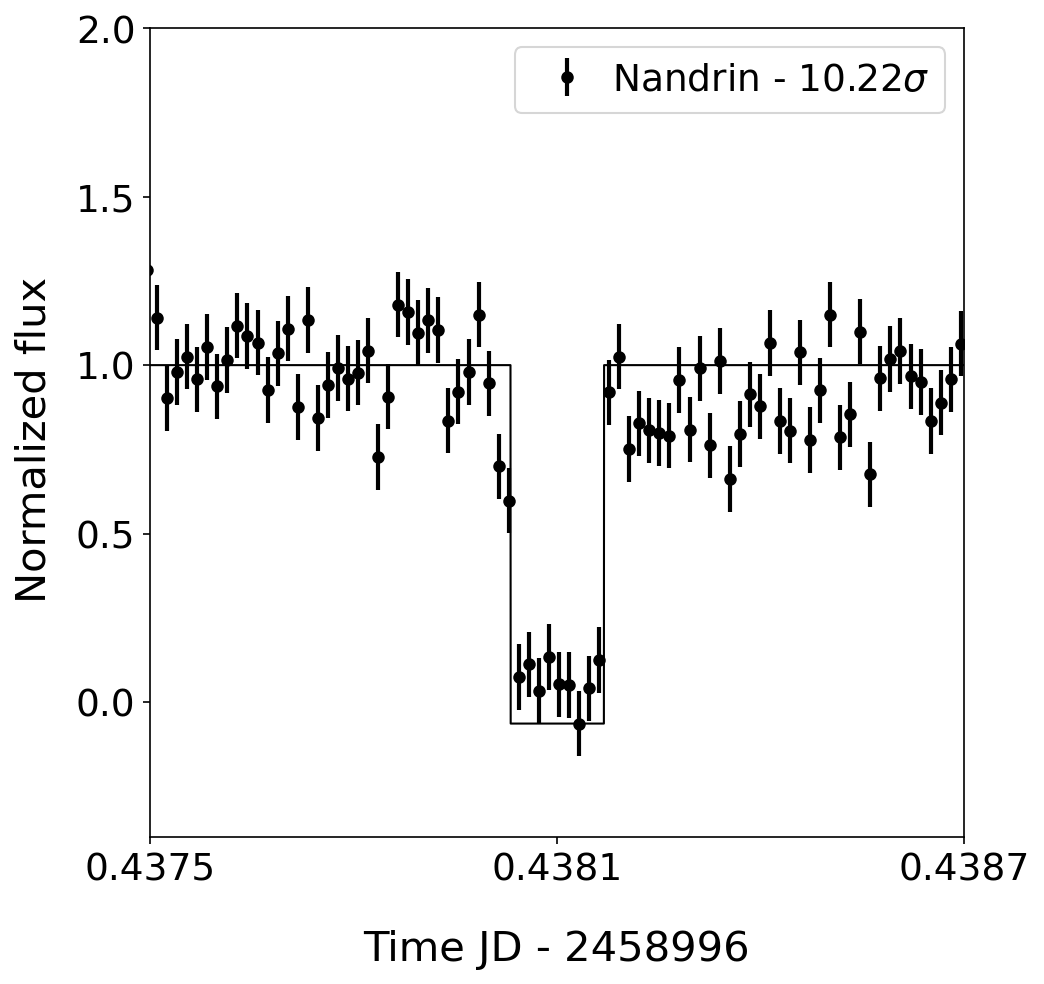}
\end{subfigure}

\caption{\label{fig:lc_appendix_stations_occA} 
Normalized flux values obtained from the six positive observations made during Occ. A on May 26, 2020. The solid line represents a square well fit to the observational data. The observing sites and the significance of the drop are indicated in the inserted labels.}
\end{figure*}

\begin{figure*}[h!]
\centering

\begin{subfigure}[t]{0.33\linewidth}
\centering
\captionsetup{justification=raggedright,singlelinecheck=false}
\caption{}
\includegraphics[width=\linewidth]{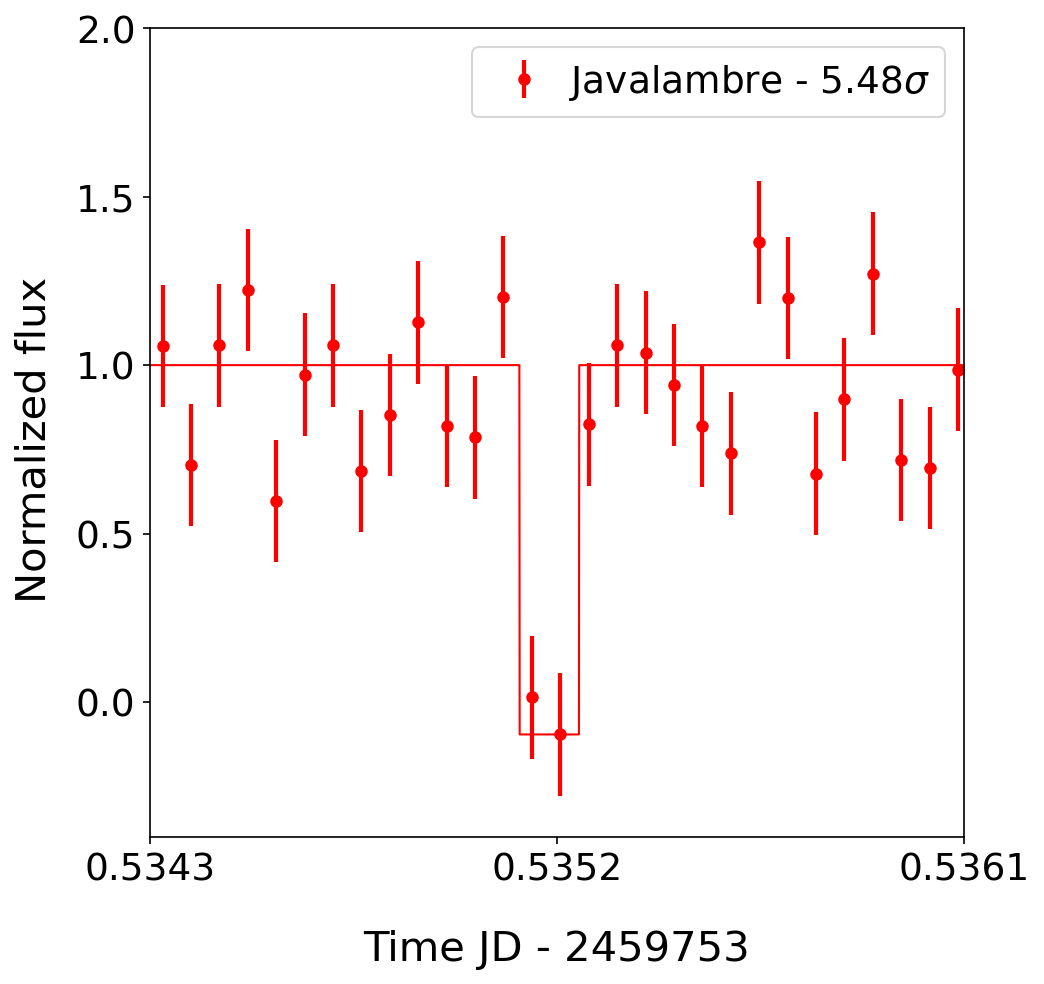}
\end{subfigure}
\hfill
\begin{subfigure}[t]{0.33\linewidth}
\centering
\captionsetup{justification=raggedright,singlelinecheck=false}
\caption{}
\includegraphics[width=\linewidth]{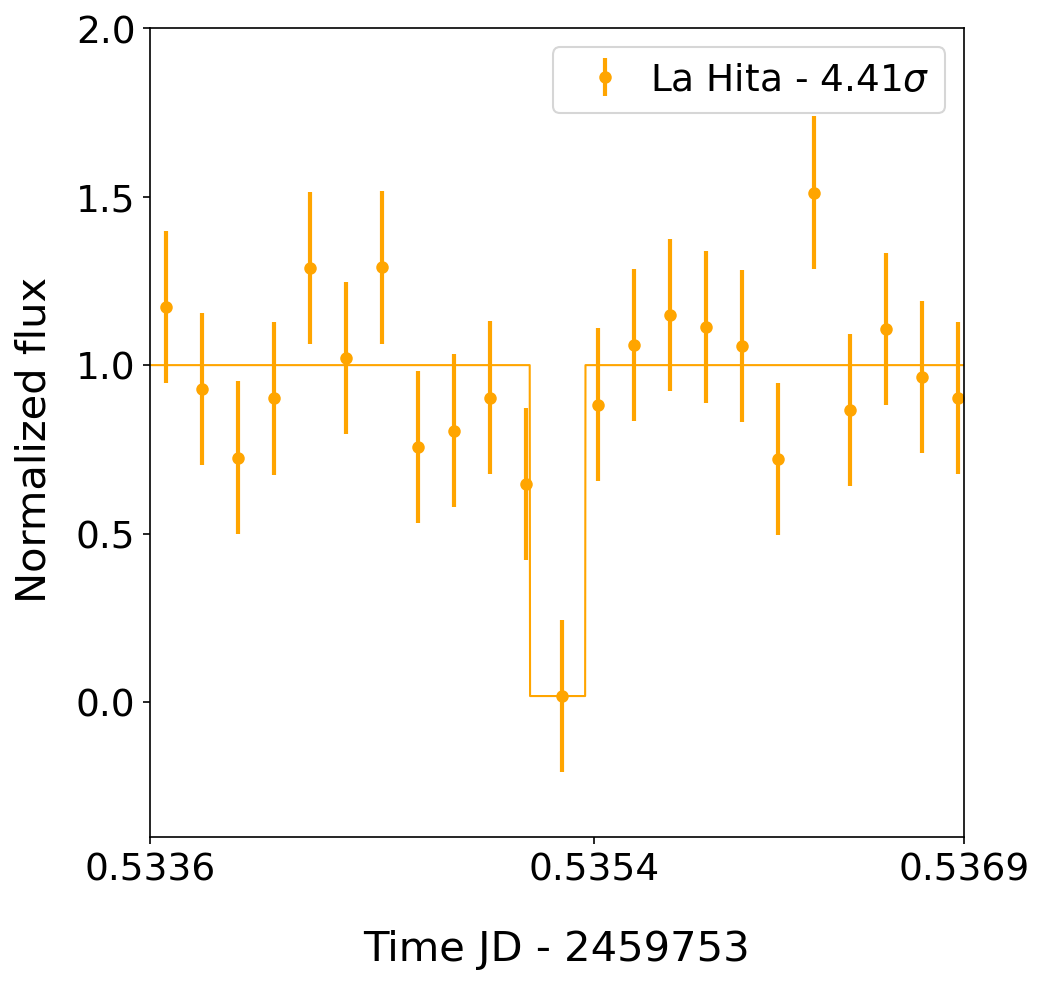}
\end{subfigure}
\medskip  
\begin{subfigure}[t]{0.33\linewidth}
\centering
\captionsetup{justification=raggedright,singlelinecheck=false}
\caption{ }
\includegraphics[width=\linewidth]{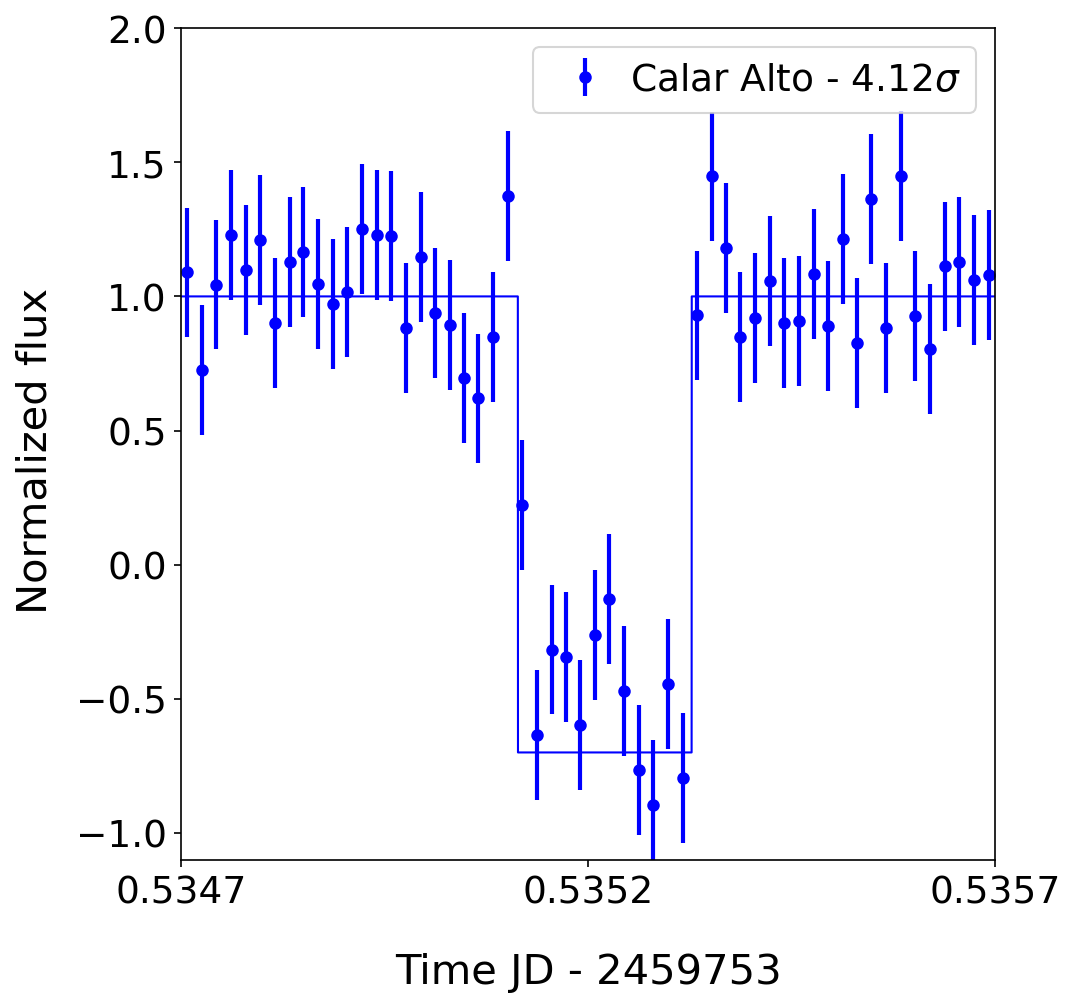}
\end{subfigure}

\caption{\label{fig:lc_appendix_stations_occB} 
Normalized flux values obtained from the three positive observations made during Occ. B on June 23, 2022. The solid line represents a square well fit to the observational data. The observing sites and the significance of the drop are indicated in the inserted labels.}
\end{figure*}

\begin{figure*}[h!]
\centering

\begin{subfigure}[t]{0.33\linewidth}
\centering
\captionsetup{justification=raggedright,singlelinecheck=false}
\caption{}
\includegraphics[width=\linewidth]{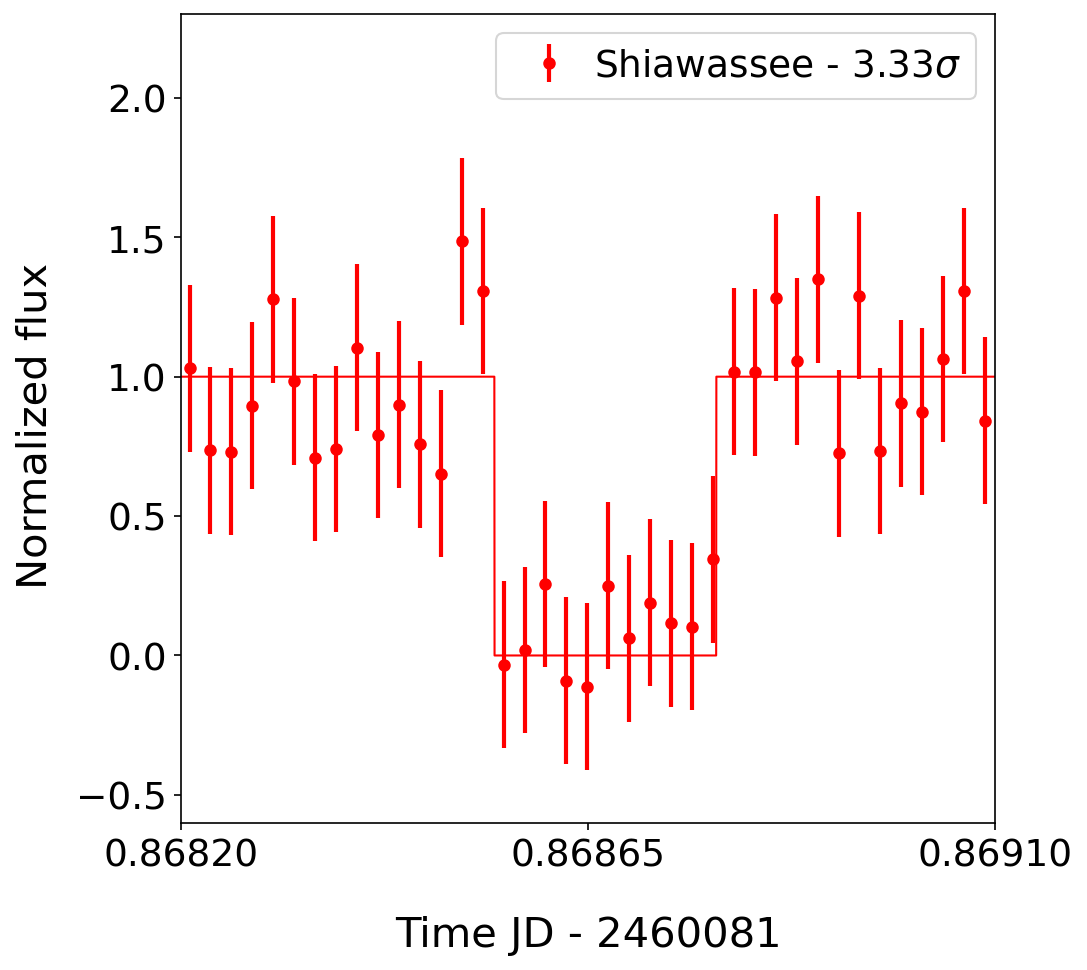}
\end{subfigure}
\hfill
\begin{subfigure}[t]{0.33\linewidth}
\centering
\captionsetup{justification=raggedright,singlelinecheck=false}
\caption{}
\includegraphics[width=\linewidth]{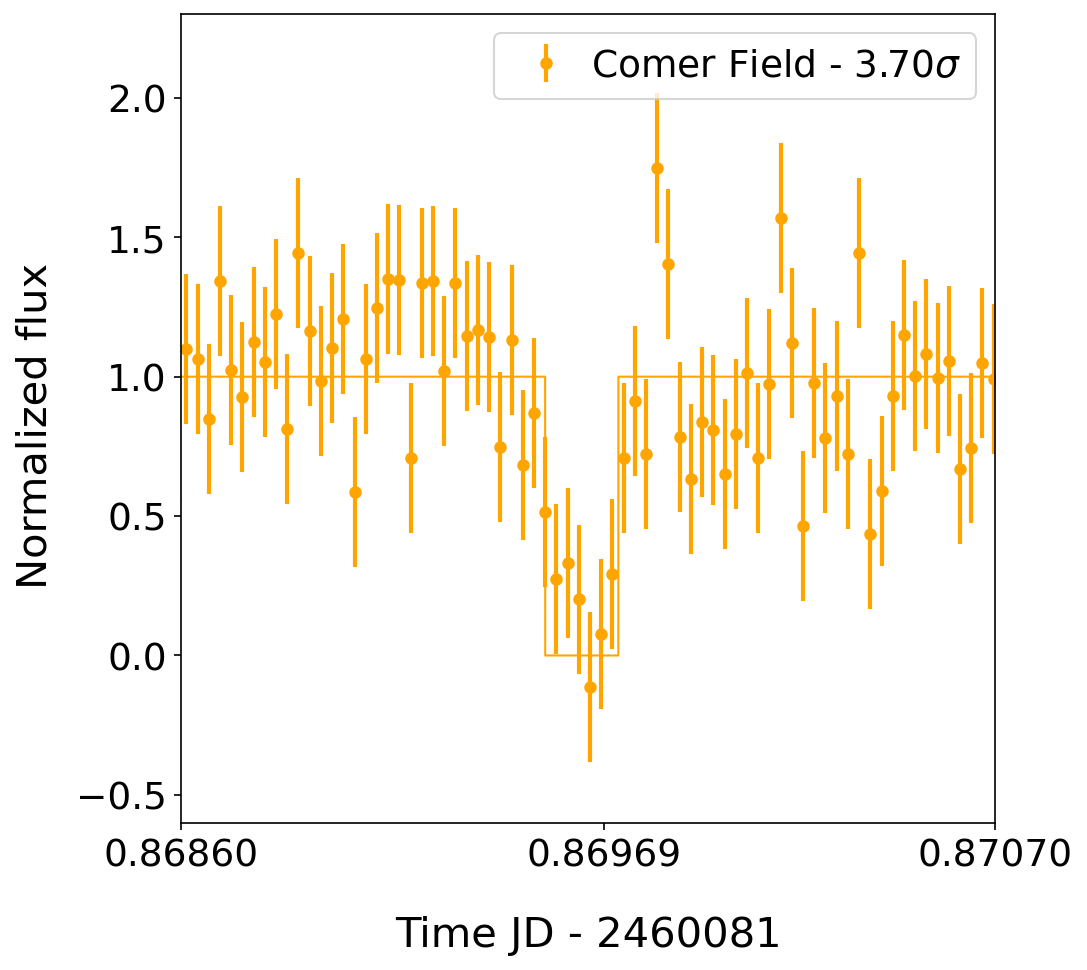}
\end{subfigure}
\hfill
\begin{subfigure}[t]{0.33\linewidth}
\centering
\captionsetup{justification=raggedright,singlelinecheck=false}
\caption{ }
\includegraphics[width=\linewidth]{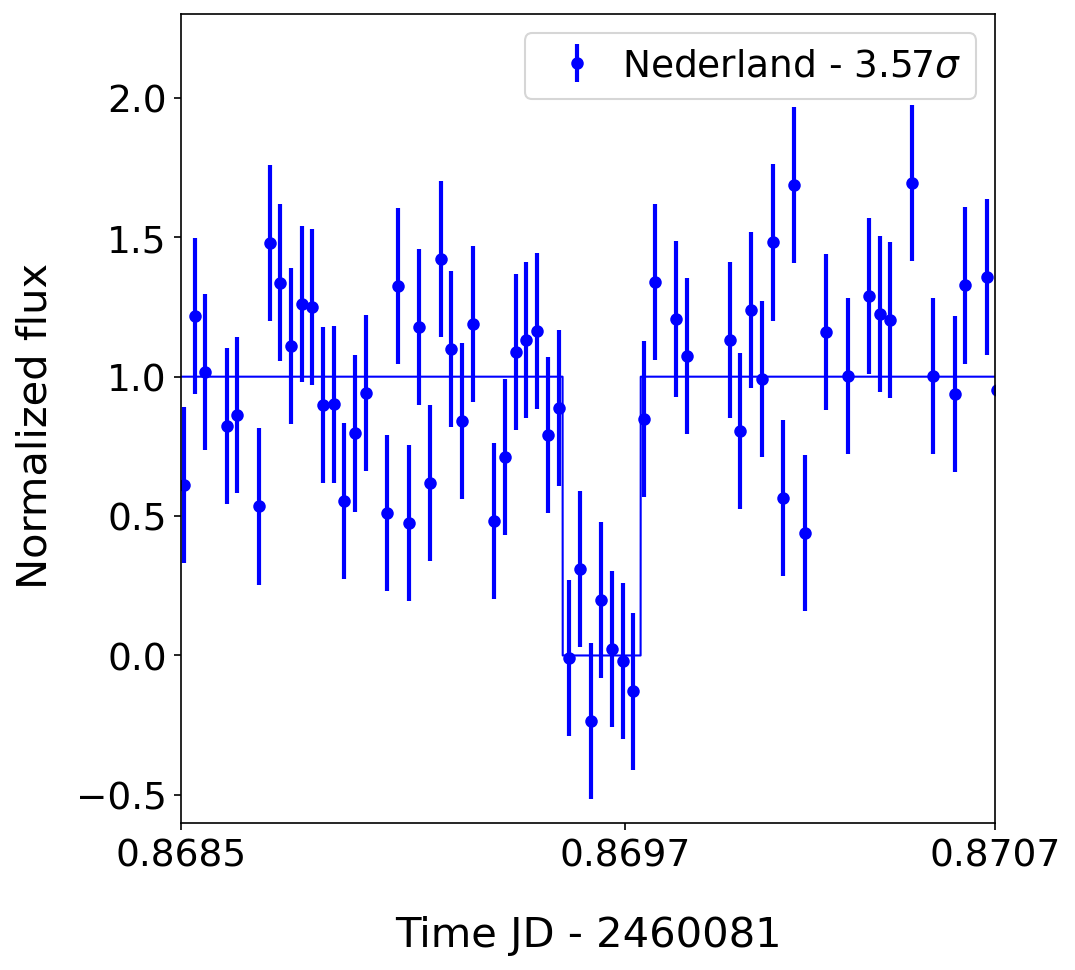}
\end{subfigure}

\caption{\label{fig:lc_appendix_stations_occC} 
Normalized flux values obtained from the three positive observations made during Occ. C on May 17, 2023. The solid line represents a square well fit to the observational data. The observing sites and the significance of the drop are indicated in the inserted labels.}
\end{figure*}

\begin{figure*}[h!]
\centering

\begin{subfigure}[t]{0.33\linewidth}
\centering
\captionsetup{justification=raggedright,singlelinecheck=false}
\caption{}
\includegraphics[width=\linewidth]{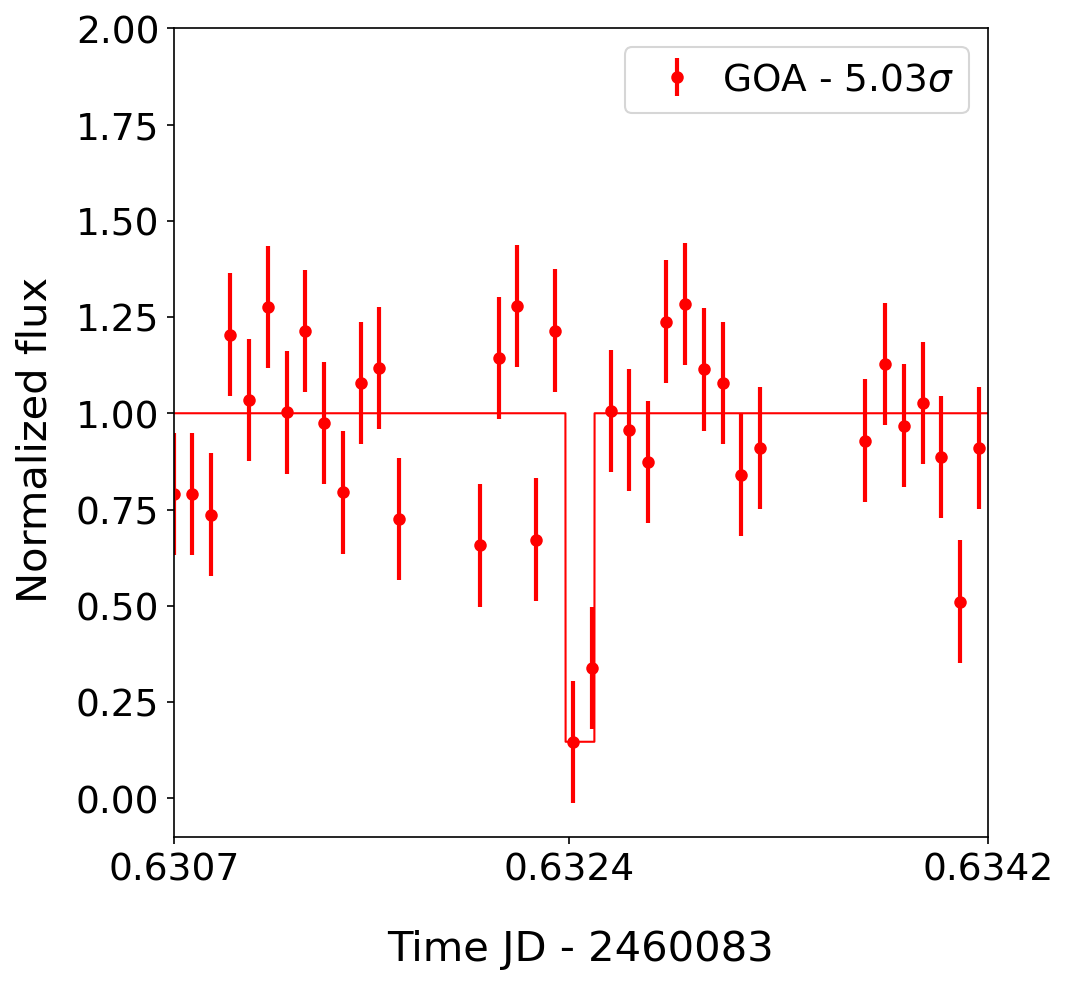}
\end{subfigure}
\begin{subfigure}[t]{0.33\linewidth}
\centering
\captionsetup{justification=raggedright,singlelinecheck=false}
\caption{}
\includegraphics[width=\linewidth]{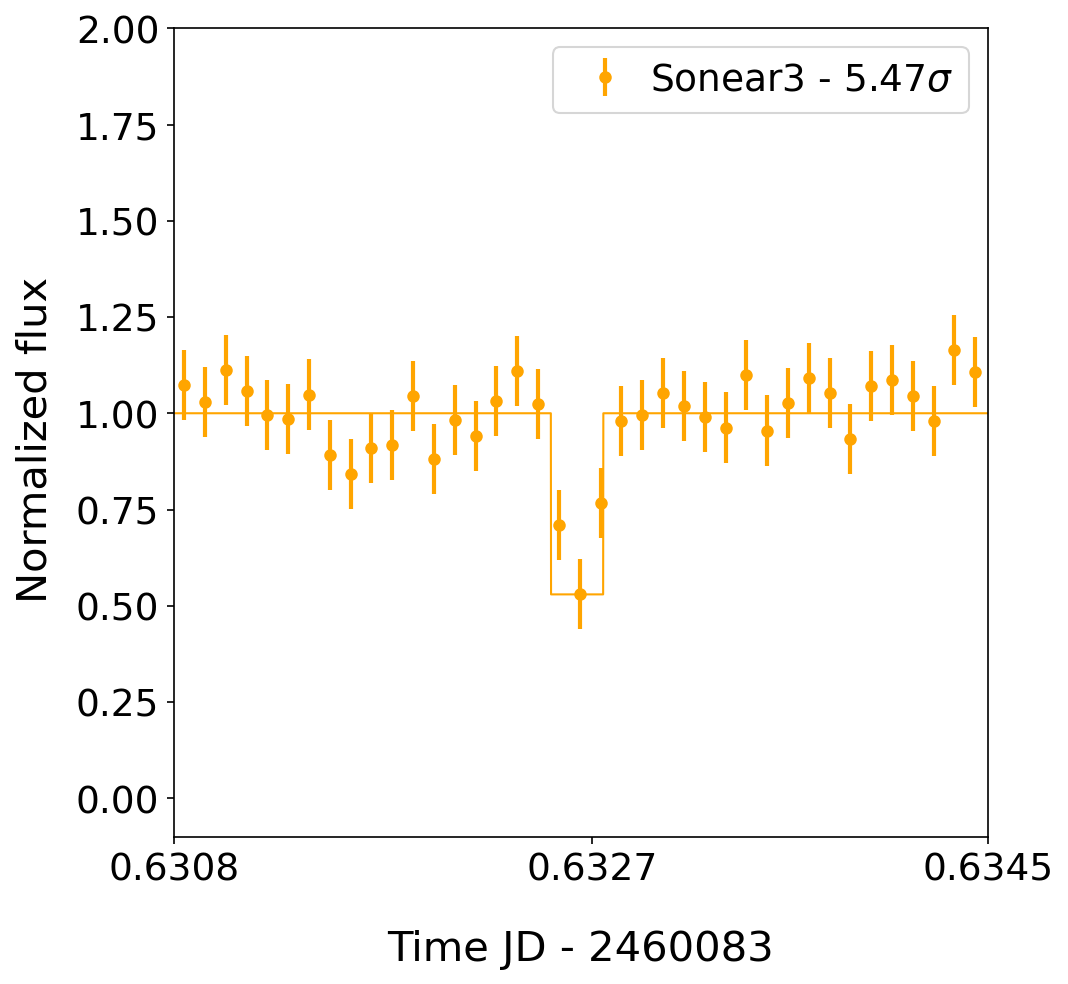}
\end{subfigure}
\begin{subfigure}[t]{0.33\linewidth}
\centering
\captionsetup{justification=raggedright,singlelinecheck=false}
\caption{}
\includegraphics[width=\linewidth]{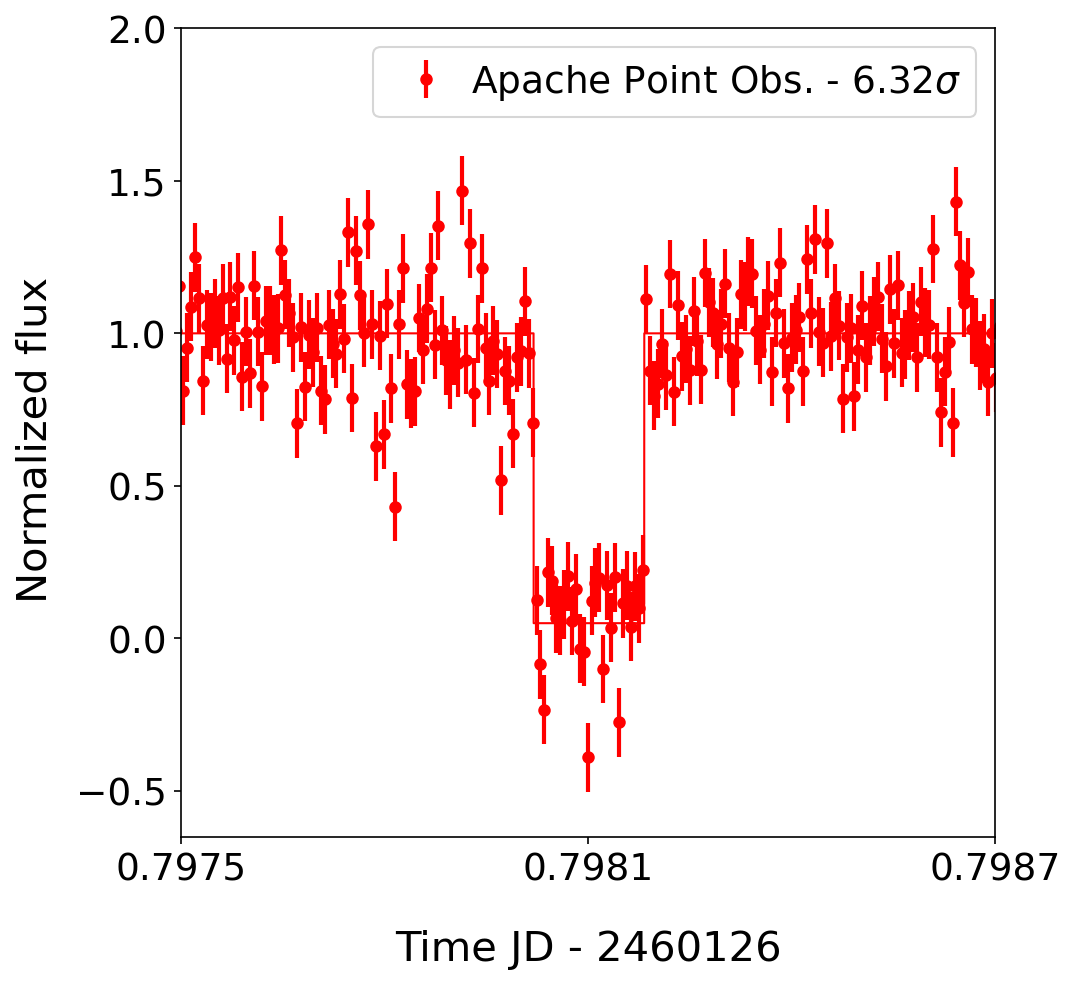}
\end{subfigure}

\caption{\label{fig:lc_appendix_stations_occD&E} 
Normalized flux values from the two positive detections, shown in (a) and (b), during Occ. D on May 19, 2023, while
(c) presents the normalized flux from the positive detection during Occ. E on July 1, 2023.
In all panels, the solid line represents a square-well fit to the observational data. The observing sites and the significance of the flux drop are indicated in the corresponding labels.}
\end{figure*}

We then conducted time series multi-aperture differential photometry in AIJ, measuring the flux of the occulted star (blended with 2002 KX14) and dividing by that of selected comparison stars within the field of view (FoV). This approach minimized systematic photometric errors arising from atmospheric fluctuations. As the event duration was only a few seconds, any flux variations due to the rotational variability of the object did not influence the resulting light curve. Aperture and sky background inner and outer radii were chosen to minimize noise in the data beyond the primary flux drop caused by the occultation. Error bars were derived from Poisson noise calculations and scaled to match the data's standard deviation \citep[see for details][]{Ortiz2020}. Finally, the ingress and egress timings, along with their uncertainties, were computed using the Python Stellar Occultation Reduction and Analysis (SORA) package \citep{Gomes2022}. The resulting light curves are shown in the Fig. \ref{fig:lc_appendix_stations_occA}, \ref{fig:lc_appendix_stations_occB}, \ref{fig:lc_appendix_stations_occC}, and \ref{fig:lc_appendix_stations_occD&E} for Occ. A, B, C, D, and E, respectively. Table \ref{table:ing_egr_times} provides the extracted ingress and egress times along with their associated uncertainties.

\begin{figure*}[h!]
\centering

\begin{subfigure}[t]{0.49\linewidth}
\centering
\captionsetup{justification=raggedright,singlelinecheck=false}
\caption{}
\includegraphics[width=\linewidth]{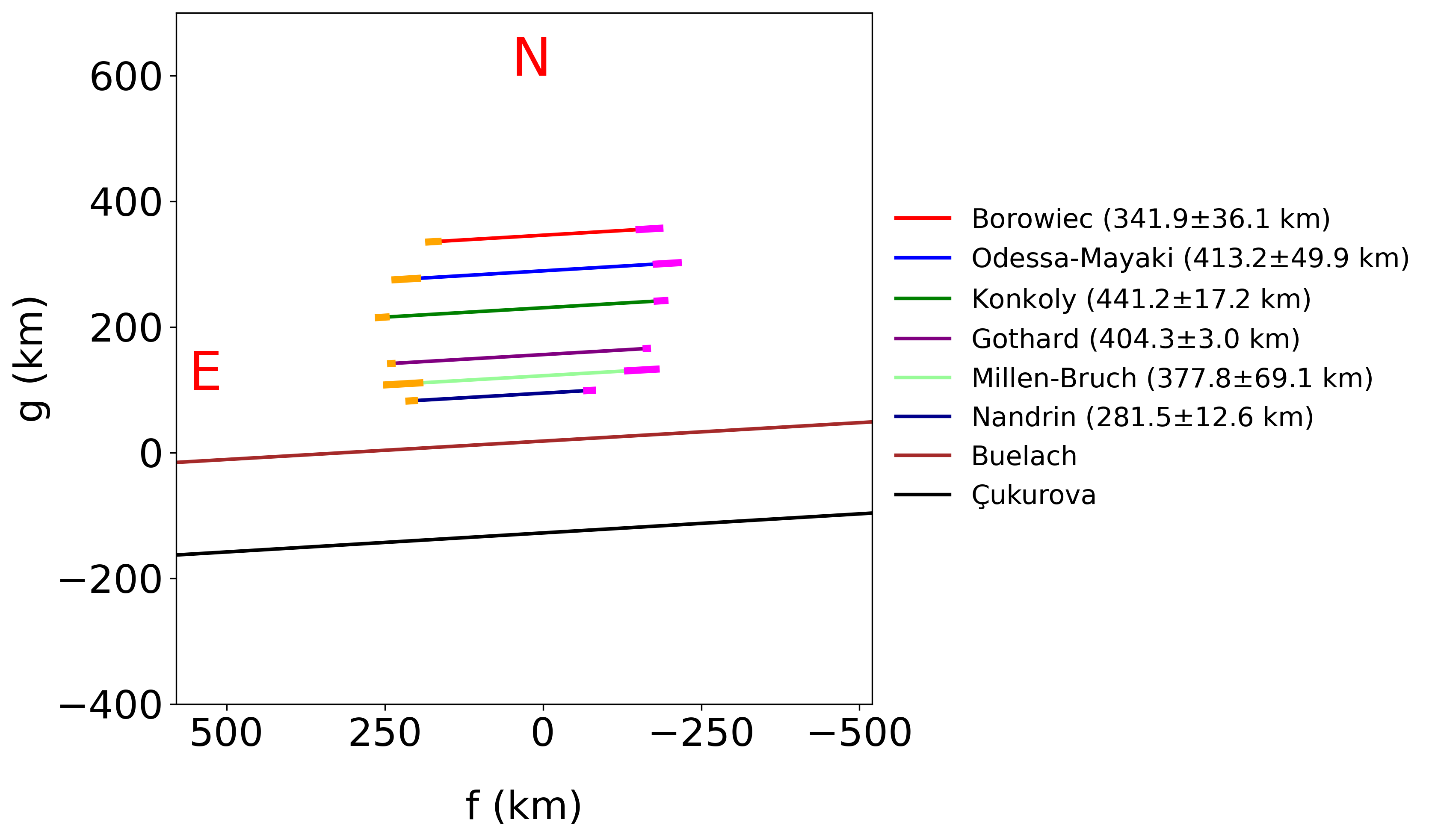}
\end{subfigure}
\hfill
\begin{subfigure}[t]{0.49\linewidth}
\centering
\captionsetup{justification=raggedright,singlelinecheck=false}
\caption{}
\includegraphics[width=\linewidth]{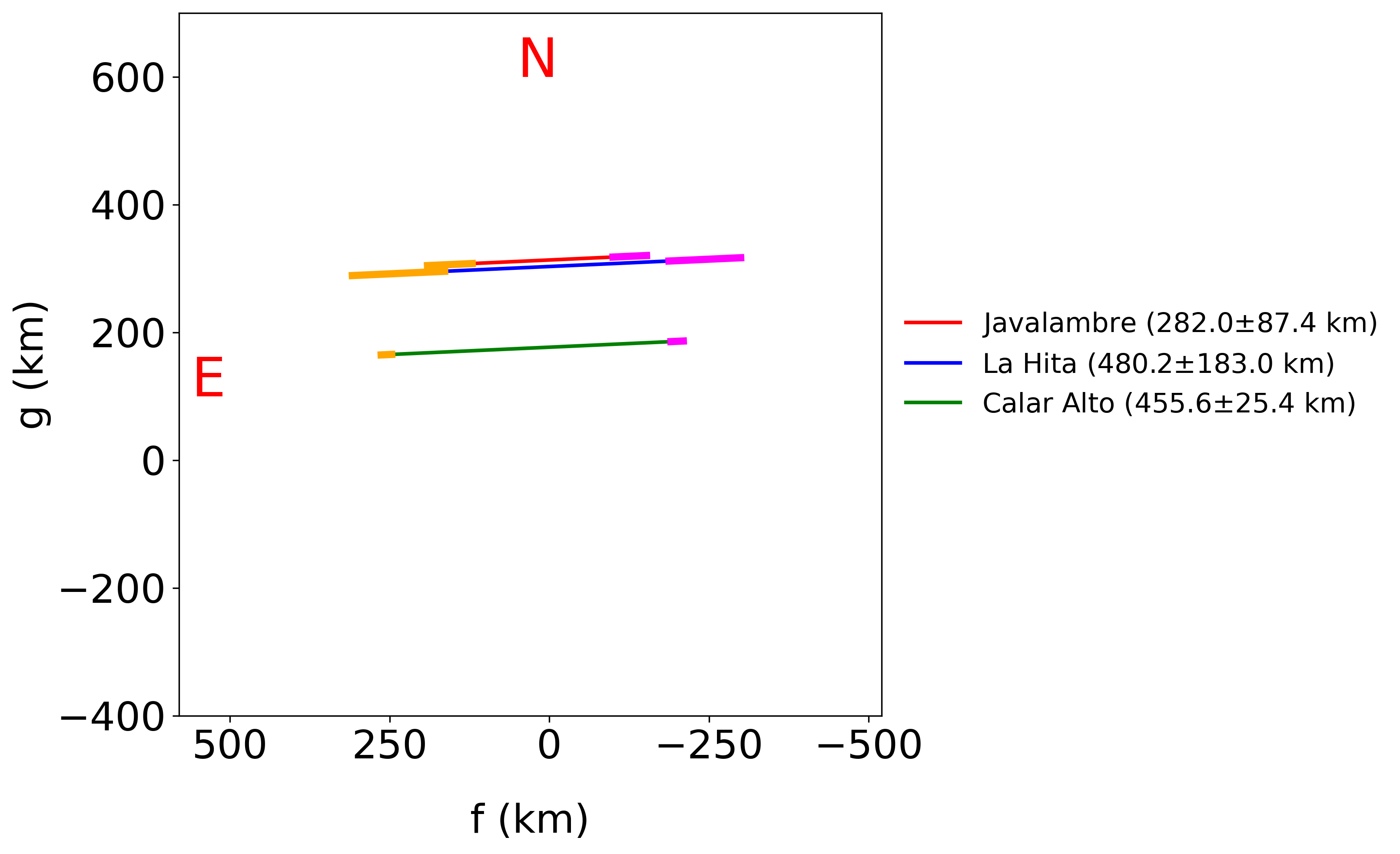}
\end{subfigure}

\medskip 

\begin{subfigure}[t]{0.49\linewidth}
\centering
\captionsetup{justification=raggedright,singlelinecheck=false}
\caption{ }
\includegraphics[width=\linewidth]{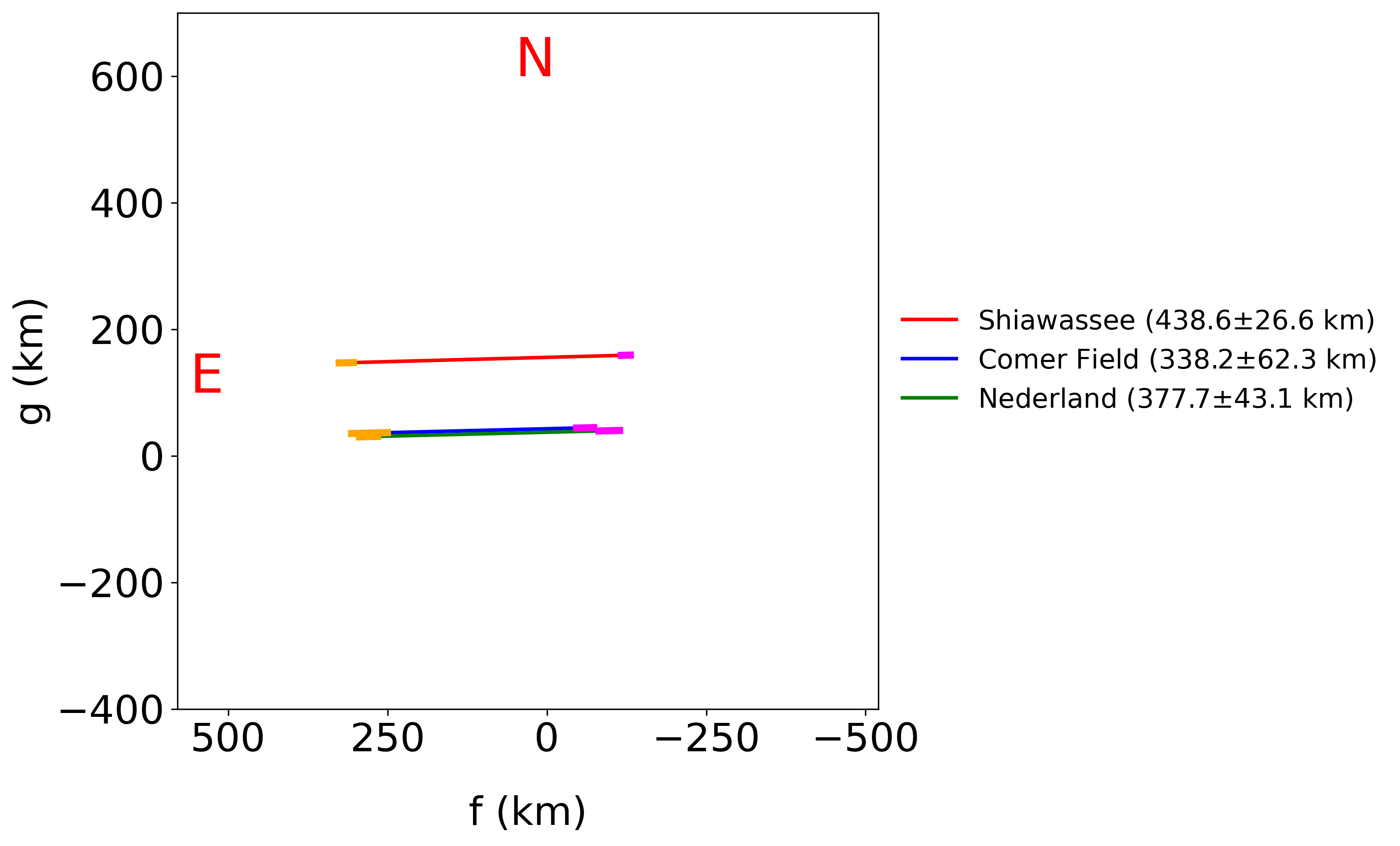}
\end{subfigure}
\hfill
\begin{subfigure}[t]{0.49\linewidth}
\centering
\captionsetup{justification=raggedright,singlelinecheck=false}
\caption{ }
\includegraphics[width=\linewidth]{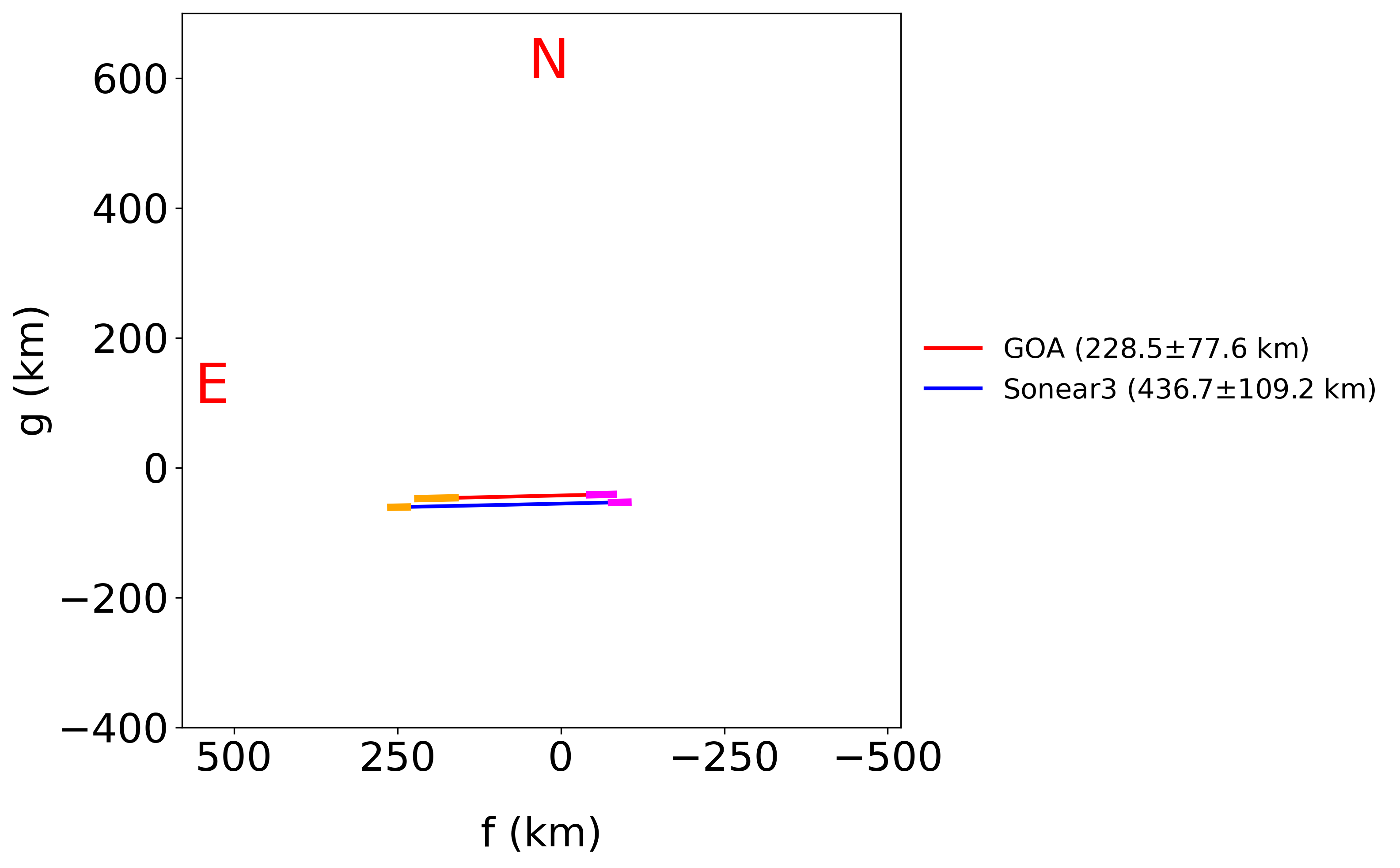}
\end{subfigure}

\medskip 

\begin{subfigure}[t]{0.49\linewidth}
\centering
\captionsetup{justification=raggedright,singlelinecheck=false}
\caption{ }
\includegraphics[width=\linewidth]{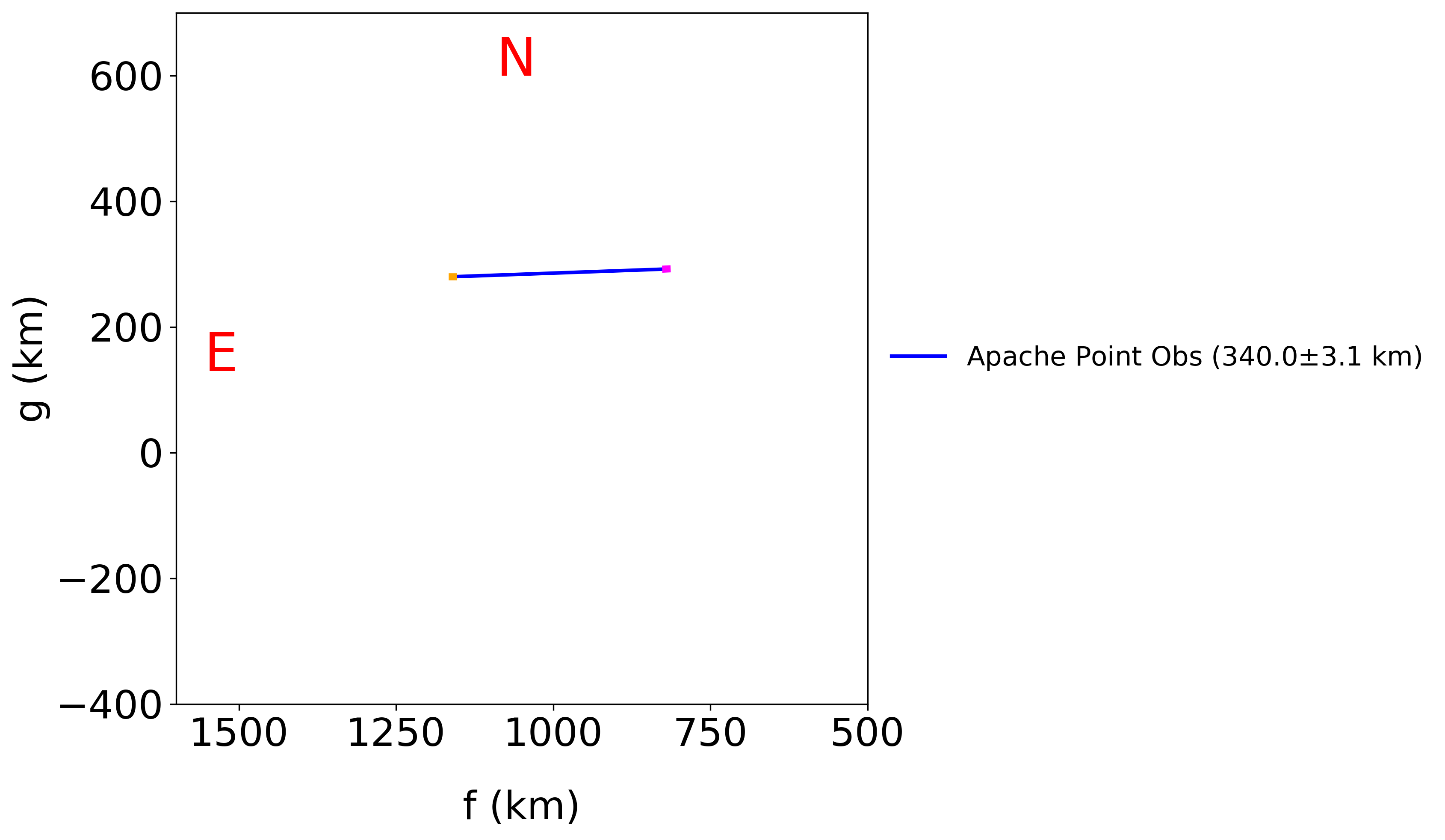}
\end{subfigure}

\caption{\label{fig:skyplane_chords} 
Chords of the stellar occultations projected on the sky plane. The ingress uncertainties are shown in pink, and the egress
uncertainties are in orange. In the legend, the chord length values in kilometers appear alongside the uncertainty error. The ephemeris source used for these plots was JPL$\#$18.}
\end{figure*}

To project the occultation chords onto the sky plane, we computed the astrometric right ascension and declination of 2002 KX14 relative to each observing site (ICRF frame, JPL$\#$18 ephemeris) at ingress and egress times shown in Table \ref{table:ing_egr_times}. This was done using the JPL Horizons online service for solar system data and ephemeris calculations. Next, we subtracted these coordinates (the chord endpoints) from those of the occulted stars, then converted this difference to kilometers in the sky plane at 2002 KX14's geocentric distance. The results for each occultation are displayed in Fig. \ref{fig:skyplane_chords}. All plots are scaled uniformly (1100 $\times$ 1100 km) for consistent comparison. We note that the centroid position of Occ. E is shifted by approximately 1000 km along the x-axis compared to the others. This suggests an issue with the star’s right ascension. Indeed, we confirm that this star has a renormalized unit weight error (RUWE) of 1.50 in the Gaia catalog. RUWE quantifies the quality of a star’s astrometric fit relative to the expected value for a single, unperturbed source, with values above 1.4 indicating potential issues in the astrometric solution. However, this did not affect our results, as the objective of this study is to determine the shape of the object, which is independent of the star’s absolute position with respect to our methodology.

There are six chords (12 endpoints) in Occ. A,  making it ideal for determining the most accurate projected ellipse. To fit the chord endpoints with an ellipse, we applied the numerically stable, non-iterative least-squares algorithm described by \cite{Radim1998}. The best-fit values are presented in Table \ref{table:occA_parameters}. To account for ingress and egress time uncertainties in the fitted ellipse, we ran a Monte Carlo simulation, generating 1000 random endpoint clones based on a Gaussian distribution centered on each chord endpoint and scaled by the respective error. We then fitted an ellipse to each clone set, determining the final ellipse parameters by averaging the results from the 1000 fits, with the standard deviation representing the error. We chose 1000 clones as this quantity showed convergence, with no further improvement in precision from adding more clones. With these values, we obtain an average diameter of $392 \pm 27$ km, lower than the thermal diameter of $ 455 \pm 27$ km estimated by \cite{Vilenius2012}. 

\begin{figure*}[h!]
        \centering
        \includegraphics[width=0.8\linewidth]{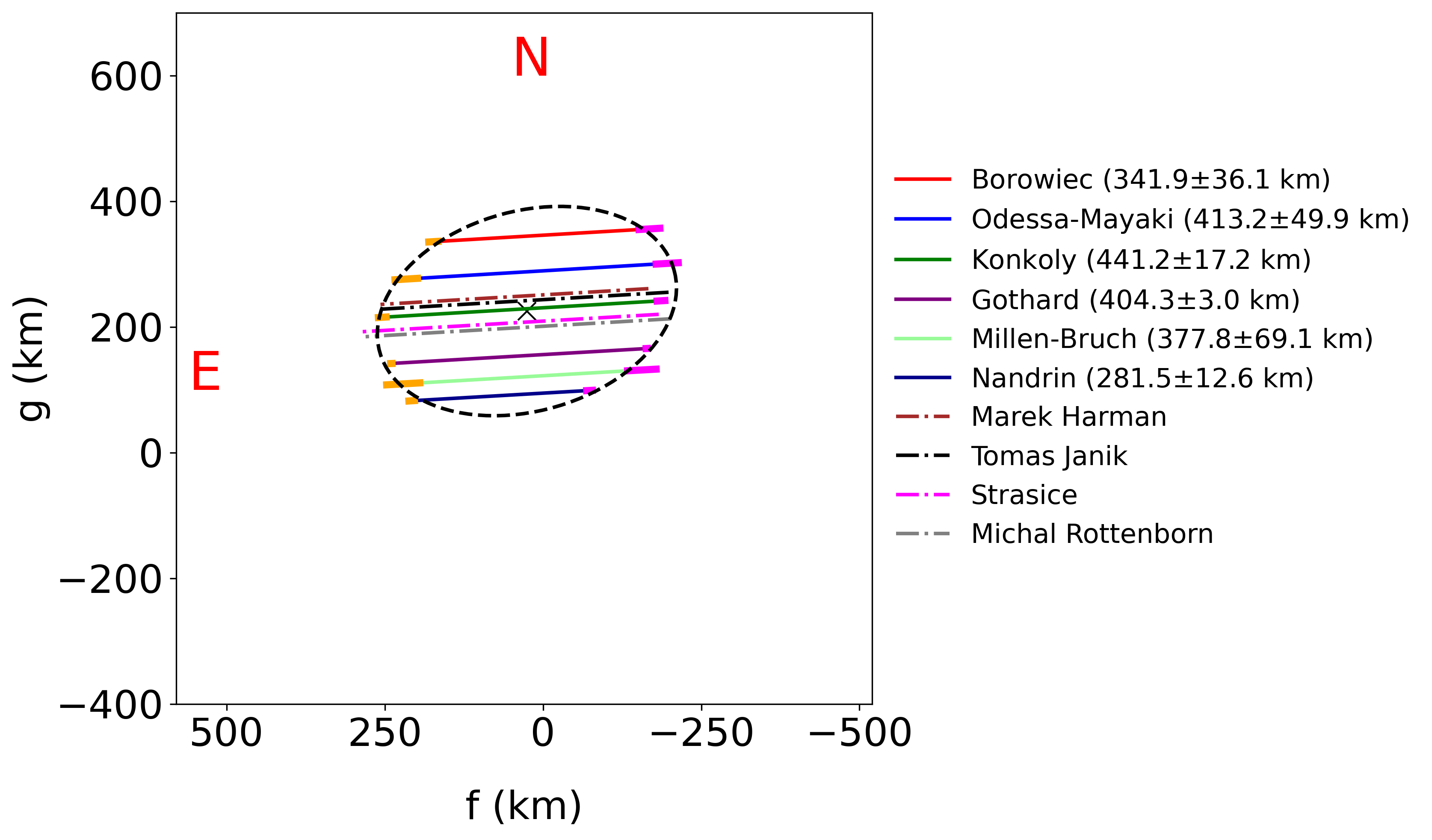}
        \caption{Six chords and the fitted ellipse for Occ. A on May 26, 2020,  shown alongside the three additional chords available in the Occult software. Since the names of the locations are unknown, the legend displays the names of the users who uploaded the data. We also include the Stracise chord, which presents technical issues. The newly incorporated chords are represented as dashed-dotted lines in this figure. We confirm that they remain consistent with the values obtained in this study.}\label{fig:occult_software}
\end{figure*}

We notice that we acknowledge the existence of three additional chords uploaded to the Occult software repository\footnote{\url{https://occultations.org/observing/software/occult/}}. However, we were unable to access the raw data necessary to ensure a consistent analysis alongside the other six chords, including the determination of ingress and egress times with their associated uncertainties using the same procedure, as well as accounting for information about the instruments or observatories. Consequently, this study includes only the six chords for which we have complete data. Nevertheless, we were able to validate our solution by incorporating these three additional chords, simply plotting the user-provided ingress and egress times (without error bars), and also including the Stracise chord, which presents technical issues. We confirm that our fitted ellipse remains consistent with these four additional chords (Figure \ref{fig:occult_software}).

\begin{table}[hbt!]
\centering
\caption{Parameters of the fitted ellipse for the chords obtained on Occ. A.}
\label{table:occA_parameters}
\renewcommand{\arraystretch}{1.5}
\begin{tabular}{l c}

        \hline
        Center coordinates ($x_{0}$, $y_{0}$) ($km$) & (25.5 $\pm$ 8.9, 225.1 $\pm$ 8.2)   \\
        \hline
        Semi-major axis, u ($km$) & 241.5 $\pm$ 11.0  \\
        \hline
        Semi-minor axis, v ($km$) & 159.0 $\pm$ 8.1 \\
        \hline
        Position angle ($^{\circ}$) & 105.8 $\pm$ 5.8 \\
        \hline

\end{tabular}

\end{table}

\begin{figure*}[h!]
\centering

\begin{subfigure}[t]{0.42\linewidth}
\centering
\captionsetup{justification=raggedright,singlelinecheck=false}
\caption{\label{fig:ellipsefitsa}}
\includegraphics[width=\linewidth]{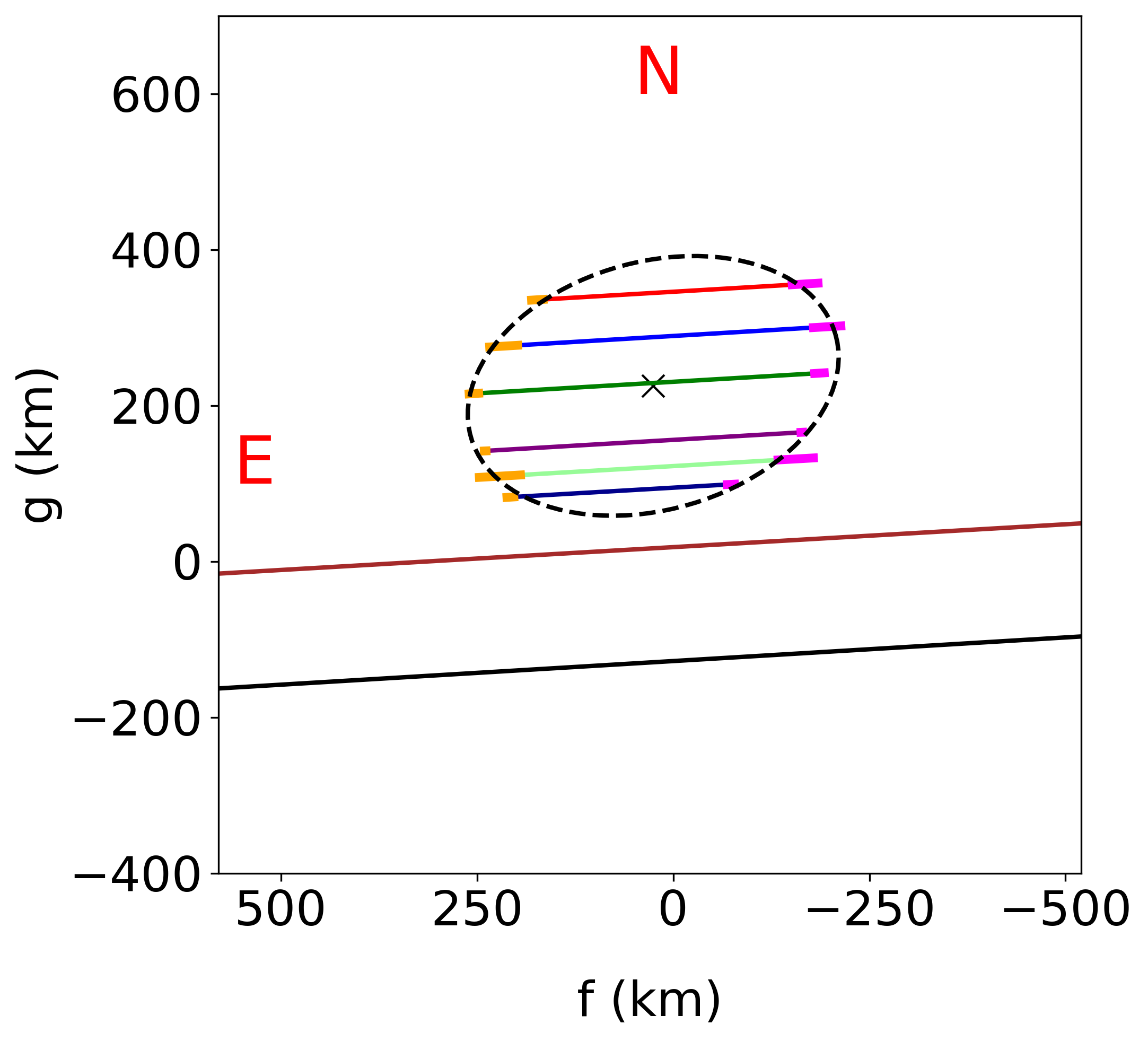}
\end{subfigure}
\hfill
\begin{subfigure}[t]{0.42\linewidth}
\centering
\captionsetup{justification=raggedright,singlelinecheck=false}
\caption{}
\includegraphics[width=\linewidth]{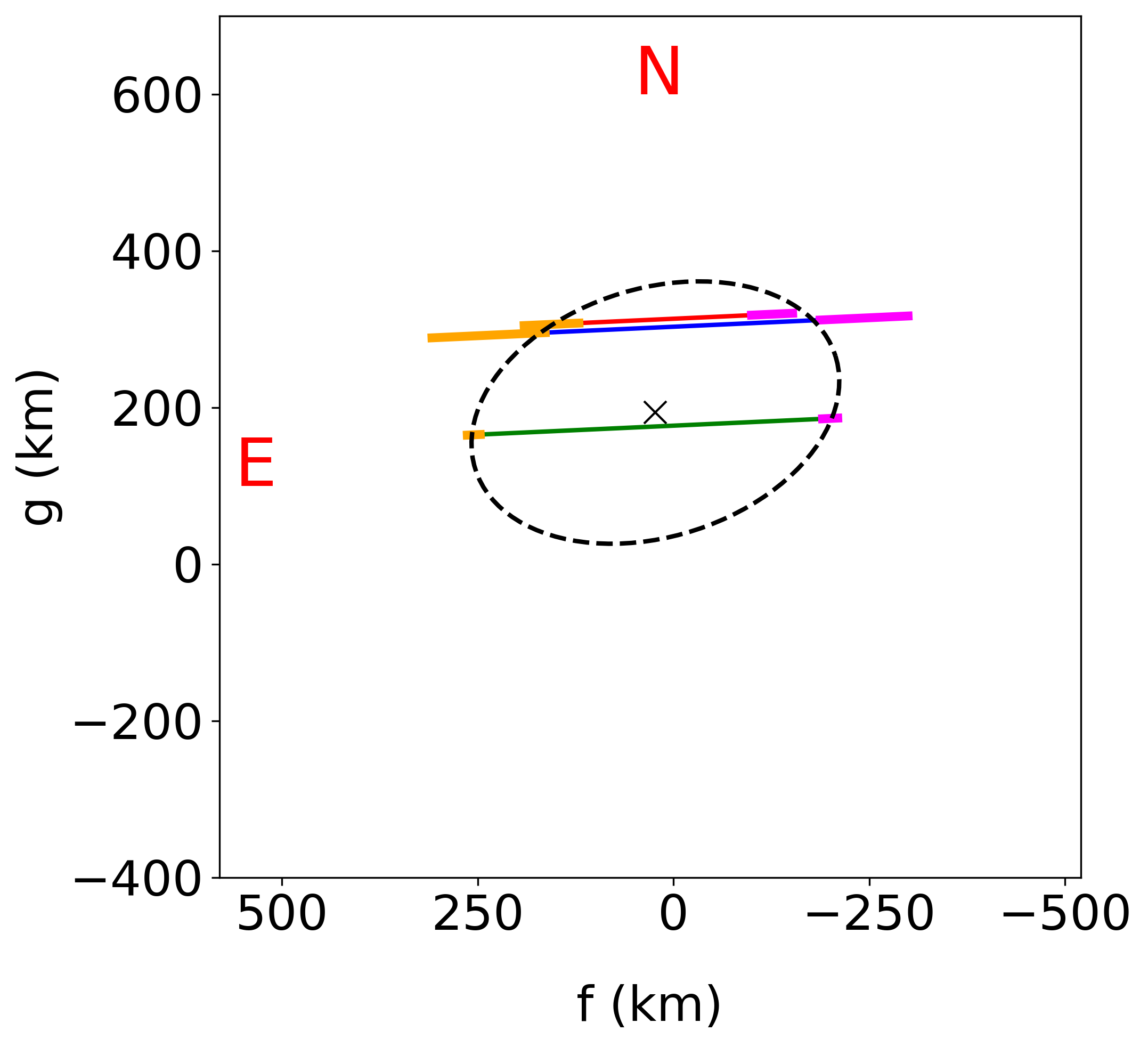}
\end{subfigure}

\medskip 

\begin{subfigure}[t]{0.42\linewidth}
\centering
\captionsetup{justification=raggedright,singlelinecheck=false}
\caption{ }
\includegraphics[width=\linewidth]{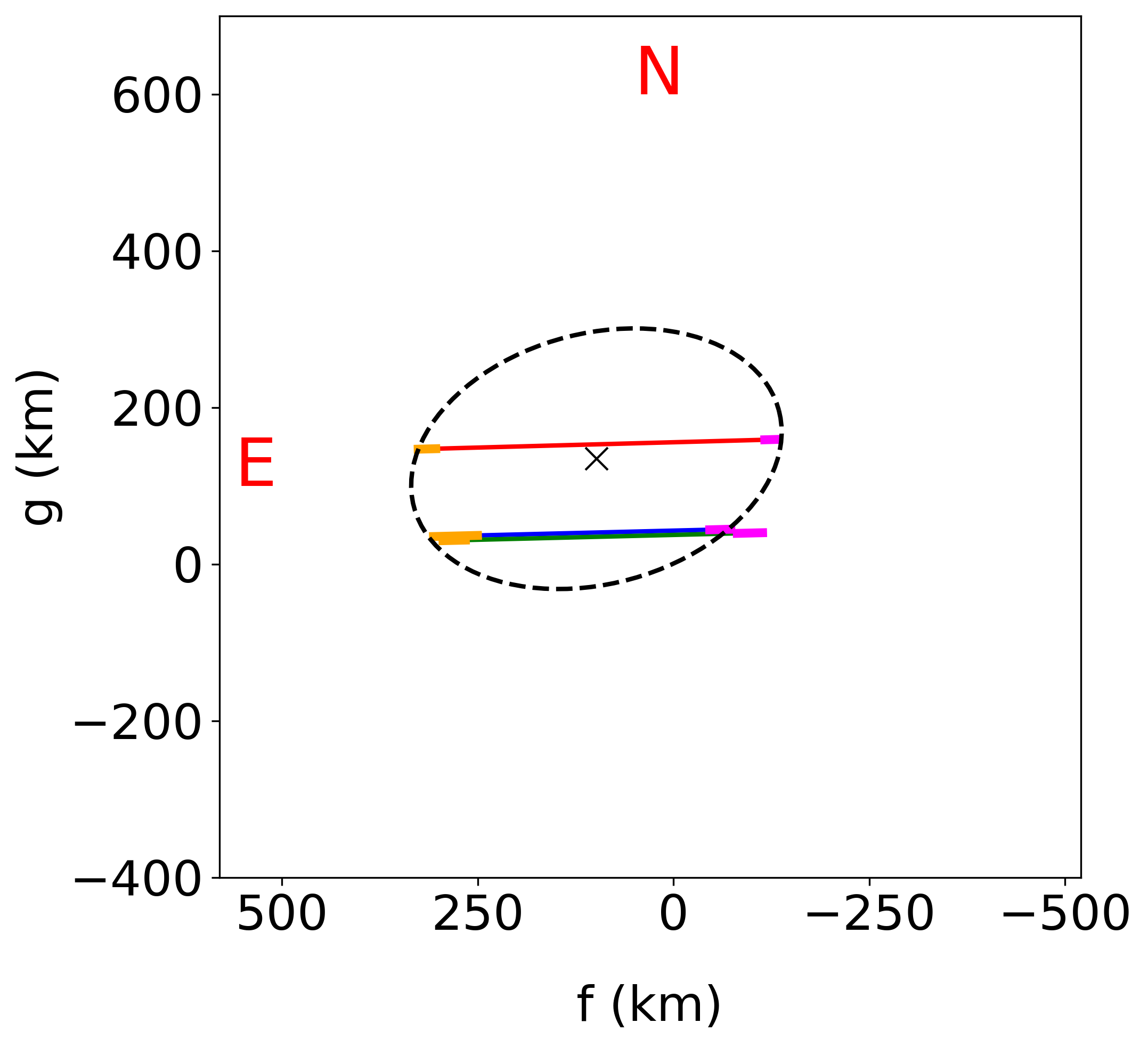}
\end{subfigure}
\hfill
\begin{subfigure}[t]{0.42\linewidth}
\centering
\captionsetup{justification=raggedright,singlelinecheck=false}
\caption{ }
\includegraphics[width=\linewidth]{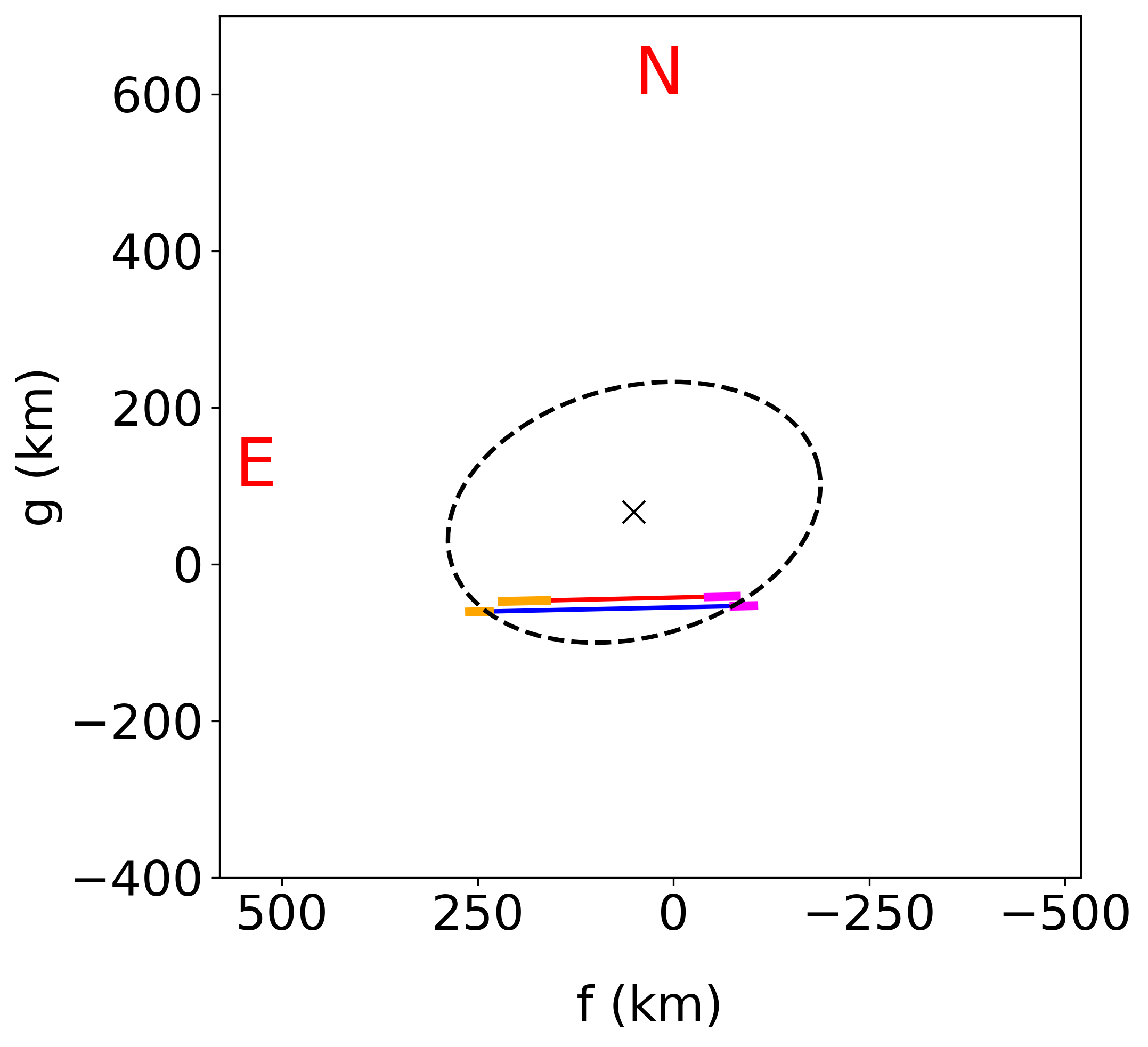}
\end{subfigure}

\medskip 

\begin{subfigure}[t]{0.42\linewidth}
\centering
\captionsetup{justification=raggedright,singlelinecheck=false}
\caption{ }
\includegraphics[width=\linewidth]{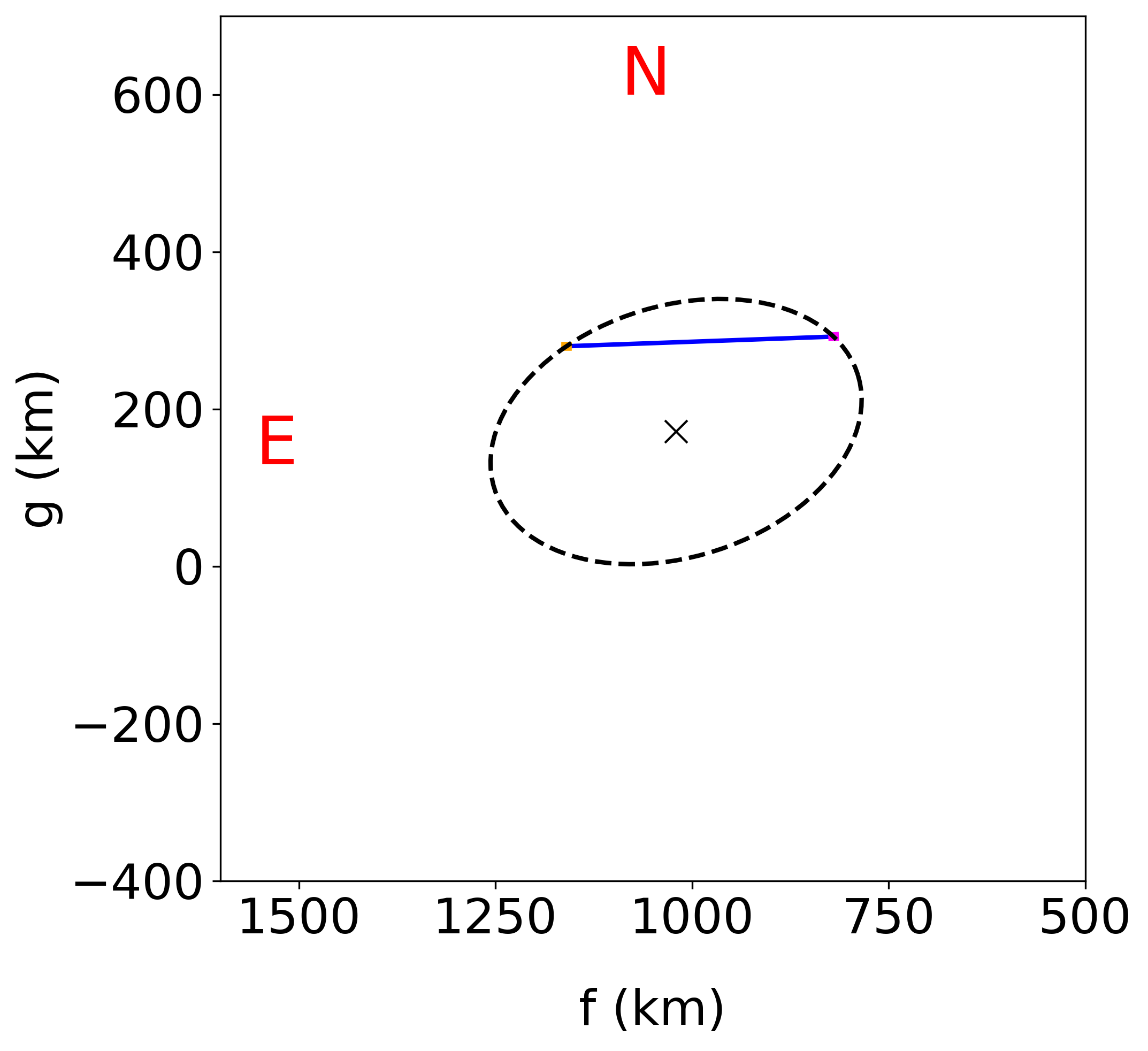}
\end{subfigure}

\caption{\label{fig:ellipsefits} 
Chords of the stellar occultations projected on the sky plane along with their fitted ellipses in dashed lines for (a) Occ. A on May 26, 2020, (b) Occ. B on June 23, 2022, (c) Occ. C on May 17, 2023, (d) Occ. D on May 19, 2023, and (e) Occ. E on July 01, 2023. The cross represents the centroid of the ellipse. The ephemeris source used for these plots was JPL$\#$18.}
\end{figure*}

Next, we determined the best-fitting ellipse for the remaining occultations. For these other cases the number of chords is significantly lower, and particularly in the case of Occ. D or E, where only one or two chords are available. Given that at least five points are required for a unique solution, we adopted a different approach. In these cases, we identified the best-fitting ellipse using a parameter grid search. To achieve this, we defined a chi-squared statistic that includes two components: a point-to-ellipse contribution ($A_{i}$), calculating the deviation of each observed point $(x_{i}, y_{i})$ from an ellipse model and taking into account the uncertainties in the coordinates, and a component that penalizes deviations of the Occ. A ellipse parameters ($B$), weighted by their respective uncertainties. The reason for defining this penalty is that, as noted by \cite{Benecchi2013}, the object V-band magnitude exhibits rotational variations below 0.05 mag, meaning that the projected shape of 2002 KX14 cannot differ significantly from the one already determined throughout the rotation cycle. Moreover, since this object is located at geocentric distances greater than 37 AU, its viewing geometry has remained nearly unchanged. For instance, over the time span covered between the first and last occultation, the phase angle ranges from 0.23 to 1.45 degrees, which is a difference of just $\sim$ 1 degree. The same reasoning applies to the rest of parameters, such as the aspect angle, regardless of the rotation pole’s orientation. Equation \ref{eq:chi2} summarizes the chi-squared statistic, as follows:

\begin{equation}\label{eq:chi2}
\chi^2 = \sum_{i=1}^{N} A_{i} + B,
\end{equation}
where:
\begin{equation}\label{eq:A}
A_{i} = \frac{\left[\left(\frac{x^{\prime}_i}{u}\right)^2 + \left(\frac{y^{\prime}_i}{v}\right)^2 - 1\right]^2}{\sigma_i^2},
\end{equation}
 with
\begin{align*}
x^{\prime}_i = (x_i - x_0)\cos(\phi) + (y_i - y_0)\sin(\phi),
 \end{align*}
\begin{align*}
y^{\prime}_i = -(x_i - x_0)\sin(\phi) + (y_i - y_0)\cos(\phi),
 \end{align*}
 and
\begin{equation}\label{eq:B}
B  =  \frac{(u - u_{A})^2}{\sigma_u^2} + \frac{(v - v_{A})^2}{\sigma_v^2} + \frac{(\phi - \phi_{A})^2}{\sigma_\phi^2}.
 \end{equation}
In Eq. \ref{eq:A}, $x^{\prime}_i$ and $y^{\prime}_i$ are the coordinates in the rotated ellipsoidal reference frame (with $x_0$ and $y_0$ being the coordinates of the center of the ellipse) and $\sigma_i = \sqrt{\sigma_{x_i}^2 + \sigma_{y_i}^2}$ is the combined uncertainty for each point. In Eq. \ref{eq:B}, $u$ and $v$ are the semi-major and semi-minor axes, respectively; $\phi$ is the position angle; $u_{A}$, $v_{A}$, and $\phi_{A}$ are the Occ. A values for $u$, $v$, and $\phi$, respectively; and $\sigma_u$, $\sigma_v$, and $\sigma_\phi$  are their corresponding uncertainties. The best grid search solutions for Occ. B, C, D, and E are summarized in Table \ref{table:elli_parameters}. The diagrams for all occultations with their best-fitting ellipses are found in Fig. \ref{fig:ellipsefits}. 

\begin{figure}[hbt!]
        \centering
        \includegraphics[width=0.9\linewidth]{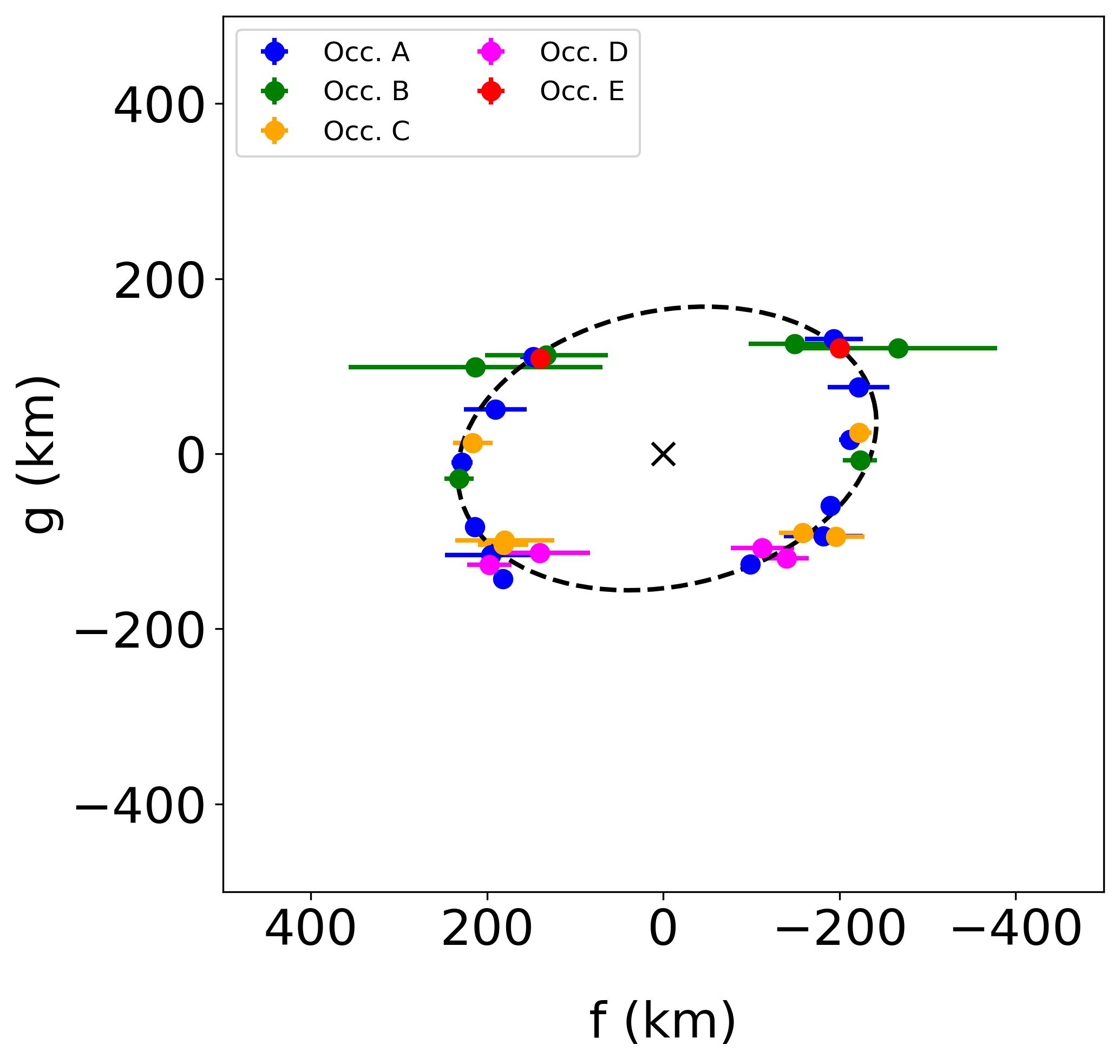}
        \caption{Chords of all the occultations centered in a common reference position, as explained in the text, along with the best fitting ellipse. This ellipse has a semi-major axis ($v$) of $241.0 \pm 7.2$ km, a semi-minor axis ($u$) of $ 157.1 \pm 5.2$ km, and an average area-equivalent diameter of $389.2 \pm 8.7$ km }
        \label{fig:all_ellipsefit}
\end{figure}

Finally, we subtracted the centroids of each ellipse ($x_0, y_0$) from their respective chords, centering them at the origin of the coordinate system. This gave us a total of 30 endpoints and we repeated the least-squares fitting using the Monte Carlo method, as described for the case of Occ. A. This new fit (Fig. \ref{fig:all_ellipsefit}) provides a more precise determination of the semi-major and semi-minor axes, $u = 241.0 \pm 7.2$ and $v = 157.1 \pm 5.2$, leading to a final value for the average area-equivalent diameter of $389.2 \pm 8.7$ km.

\subsection{Geometric albedo}

Unfortunately, the rotational period of 2002 KX14 remains unknown. However, through this new average diameter in the visible, along with the most accurate visual magnitude found in the literature \citep[$H_{V}$ = 4.978 $\pm$ 0.017;][]{Alvaro2016}, it is possible to calculate its geometric albedo using Eq. \ref{eq:geometric} \citep{Russell1916}, where $C$ = 1330 ± 18 km is a constant \citep{Masiero2021}. 

\begin{equation}\label{eq:geometric}
\sqrt{p_{V}} = \frac{C}{D}10^{-H_{V}/5}.
\end{equation}

This yields a value of $p_{V} = 11.9 \pm 0.7\%$.

\section{Conclusions and future works}\label{sec:conclusions}

This study presents the analysis of five stellar occultations by the TNO (119951) 2002 KX14 observed between 2020 and 2023. These events yielded 15 positive chords, enabling precise determination of the object's size, shape, and albedo. Using these occultation data, we derived an average area-equivalent diameter of  $389.2 \pm 8.7$ km, compatible with the $365\substack{+20 \\ -21 }$ obtained by \cite{Alvaro2014}, and lower than that estimated by \cite{Vilenius2012}. The geometric albedo is estimated at $11.9 \pm 0.7\%$, higher than that obtained through thermal observations but within error bars at 3$\sigma$.

Objects with a variability lower than 0.15 mag are possibly Maclaurin spheroids (i.e., oblate bodies with two equal axes larger than the third), where this low variability is likely due to the presence of large-scale albedo features on their surface \citep{Thirouin2010}. Thus, given the minimal rotational variability of 2002 KX14, we can reasonably assume that the observed variations are due to albedo features. Since the observed axis ratio ($u/v$) is 1.53, if we assume a spheroidal shape and adopt a rotation period of 7 hours, which is consistent with the typical values estimated for trans-Neptunian objects \citep{Duffard2009}, we derive a density of approximately 900 kg m$^{-3}$ using the Chandrasekhar formalism \citep{Chandrasekhar1987}. That said, it is important to note that this density estimate relies on several assumptions that are not directly supported by our dataset, such as a rotation period of 7 hours, although significantly longer periods are equally plausible.

The refined measurements of 2002 KX14's physical properties demonstrate the effectiveness of stellar occultations in characterizing distant objects in the outer solar system. Currently, the projected shapes of only 13 other TNOs have been determined; among them, 2002 KX14 will be the first cold classical object (although, as explained in Section \ref{sec:intro}, it is possible that it could belong to the hot population). Either way, regardless of its classification, these results contribute to the growing body of knowledge on the trans-Neptunian region.

In recent years, 2002 KX14 has been passing through the dense star fields of the Milky Way, complicating the acquisition of high-quality light curves needed for determining its rotation period. This difficulty is further compounded by its low rotational amplitude, which remains below 0.05 mag. As the object moves away from this crowded region, opportunities will arise to collect high-precision photometric data. Combining such data with the results presented here will enable the determination of a rotational period and a more comprehensive understanding of 2002 KX14’s physical properties. Additionally, further occultation campaigns may refine the shape and albedo estimated in this work, particularly if more multi-chord events are observed. These efforts could significantly reduce uncertainties in the current models, allowing for the generation of more accurate shape models and potentially determining the object’s spin axis orientation. Future studies will also integrate these results with thermal modeling and spectroscopy to better constrain the object’s surface composition and thermal properties, offering a deeper insight into its evolutionary history.

\begin{acknowledgements}
We express our gratitude to observers Marek Harman, Tomas Janik, and Michal Rottenborn for sharing their analysis and contributing it to the Occult software repository. We thank Annie Peck for generously donating observing time for Occ. E analyzed in this study.  We thank Riley DeColibus and E. Fernandez for collaborating in the observations. J. L. Rizos, J.L. Ortiz, N. Morales, P. Santos-Sanz, R. Leiva, M. Kretlow, R. Morales, A. Alvarez-Candal, R. Duffard and J. M. Gómez-Limón acknowledge financial support from the Severo Ochoa grant CEX2021-001131-S funded by MCIN/AEI/10.13039/501100011033. J. L. Rizos acknowledges support from the Ministry of Science and Innovation under the funding of the European Union NextGeneration EU/PRTR. F.L.Rommel acknowledges CNPq, Brazil grant 103096/2023-0, Florida Space Institute’s Space Research Initiative, 379 and the University of Central Florida’s Preeminent Postdoctoral Program (P3). Part of this work was supported by the Spanish projects PID2020-112789GB-I00 from AEI and Proyecto de Excelencia de la Junta de Andalucía PY20-01309. P. Santos-Sanz acknowledges financial support from the Spanish I+D+i project PID2022-139555NB-I00 (TNO-JWST) funded by MCIN/AEI/10.13039/501100011033. A. Alvarez-Candal acknowledges financial support from the Spanish project PID2023-153123NB-I00, funded by MCIN/AEI. GyMSz thanks the support by the SNN-147362 and the GINOP-2.3.2-15-2016-00003 grants of the Hungarian Research, Development and Innovation Office (NKFIH). This work has made use of data from the European Space Agency (ESA) mission \textit{Gaia}, processed by the Gaia Data Processing and Analysis Consortium (DPAC). Funding for the DPAC has been provided by national institutions, in particular the institutions participating in the Gaia Multilateral Agreement. This work is partly based on observations collected at the Centro Astronómico Hispano en Andalucía (CAHA) at Calar Alto, operated jointly by Junta de Andalucía and Consejo Superior de Investigaciones Científicas (CSIC). This research is also partially based on observations carried out at the Observatorio de Sierra Nevada (OSN) operated by Instituto de Astrofísica de Andalucía (IAA-CSIC). This work is also partly based on observations made with the Tx40 telescope at the Observatorio Astrofísico de Javalambre in Teruel, a Spanish Infraestructura Cientifico-Técnica Singular (ICTS) owned, managed, and operated by the Centro de Estudios de Física del Cosmos de Aragón (CEFCA). Tx40 is funded by the Fondos de Inversiones de Teruel (FITE). This work is also partly based on observations obtained with the Apache Point Observatory 3.5-meter telescope, which is owned and operated by the Astrophysical Research Consortium.

\end{acknowledgements}

\bibliography{main}

\begin{appendix}
\onecolumn

\section{Hertzsprung–Russell diagram of occulted stars and predicted occultation shadow path }

Hertzsprung–Russell diagram showing stars from the HYG star database archive\footnote{\url{https://github.com/astronexus/HYG-Database?tab=readme-ov-file}}, which combines data from the HIPPARCOS, Yale Bright Star, and Gliese (nearby star) catalogs. The occulted stars analyzed in Section \ref{sec:observations} are plotted and are all giant stars. To retrieve the absolute magnitude and temperature of the occulted stars, we used data from the GAIA catalog. The absolute magnitude was calculated using the star’s apparent G magnitude, which was then converted to absolute magnitude via parallax.

\begin{table*}[t]
  \centering
  \rotatebox{90}{%
    \begin{minipage}{1\textheight}
      \def\arraystretch{1.2}
      \small\centering
      \setlength{\tabcolsep}{1.0mm}
      \scriptsize
      \centering
      \caption{Observatories, instrumentation, and observers.}\label{table:observatories}  
      \begin{tabular}{l p{4cm} p{2cm} p{2cm} p{1cm} p{2cm} p{2.5cm} p{1cm} p{1cm} p{1.8cm} p{1.8cm} p{2cm}}

        \textbf{Date} & \textbf{Observatory}  & \textbf{Longitude} & \textbf{Latitude} & \textbf{Altitute (m)} & \textbf{Aperture diameter} & \textbf{Detector} & \textbf{Exposure time (s)} & \textbf{Cicle time (s)} & \textbf{Synchronization} & \textbf{Occultation} & \textbf{Observers}\\
    
          \hline
         
         26 May, 2020 & Borowiec (Poland)  & 17$^\circ$04'30.8"E& 52$^\circ$16'36.8"N & 83.0 & 400 mm & SBIG ST7  & 3.0 & 5.0& NTP& Positive & \makecell{A. Marciniak}\\
        
         \hline
        
         26 May, 2020 & Odesa-Mayaki Observatory (Ukraine) & 30$^\circ$16'18.1"E & 46$^\circ$23'50.2"N & 23.28 & 800 mm & CCD FLI ML09000 & 0.5& 1.9 & NTP & Positive & \makecell{V. Kashuba \\ N. Koshkin \\ S. Kashuba }  \\
         
        \hline
        
        26 May, 2020 & Konkoly Observatory (Hungary)       & 19$^\circ$53'41.7"E & 47$^\circ$55'06.0"N  & 477.0 & 800 mm & ASI178MM        & 1.0 & 1.0 & NTP & Positive &    \makecell{A. Pal}    \\

        \hline

        26 May, 2020 & Strasice (Czech Republic)      & 13$^\circ$44'56.3"E & 49$^\circ$44'35.7"N  & 538.0 & 200 mm & WATEC120N+ & 5.12 & 5.12 & NTP & Positive &  \makecell{J. Kubanek}    \\

        \hline
        
        26 May, 2020 & Gothard Observatory (Hungary) & 16$^\circ$36'15.7"E & 47$^\circ$15'29.8"N & 200.0 & 800 mm & Photometrics Prime 9bB & 0.7 & 0.7 & NTP & Positive & \makecell{G. M. Szabó \\ A. Derekas \\ L. Szigeti }  \\

        \hline
        
        26 May, 2020 & Millen-Bruch (Germany) & 05$^\circ$53'47.2"E & 51$^\circ$01'38.8"N & 56.0 & 250 mm & WAT910HX & 0.7 & 0.7 & GPS & Positive &   \makecell{C. Ellington}  \\

        \hline

        26 May, 2020 & Nandrin (Belgium) & 05$^\circ$26'29.5"E & 50$^\circ$31'24.8"N & 260.9 & 406 mm & WATEC910HX/RC & 1.3 & 1.3 & GPS & Positive & \makecell{ O. Schreurs }  \\
     
        \hline
        
        26 May, 2020 & Buelach (Switzerland) & 08$^\circ$34'14.3"E & 47$^\circ$31'10.4"N & 550.0 & 500 mm & IMX174LLJ & 0.5 & 1.3 & GPS & Negative &  \makecell{A. Schweizer}  \\

        \hline

        26 May, 2020 & Çukurova (Turkey) & 35$^\circ$21'12.6"E & 37$^\circ$03'15.6"N & 260.9 & 500 mm & Unknown & 9.0 & 10.0 & Unknown & Negative &   \makecell{M. Tekes}  \\

        \hline   

        26 May, 2020 & Tubitak Observatory (Turkey) & 30$^\circ$20'07.9"E & 36$^\circ$49'17.1"N & 2538.7 & 1000 mm & Apogee Alta U47 & 2.0 & 1.0 & GPS & S/N low & \makecell{ Y. Kilic \\ O. Erece \\ M. Kaplan }  \\

        \hline  

        26 May, 2020 & Marseilles (France) & 05$^\circ$23'42.3"E & 43$^\circ$18'21.1"N & 68.9 & 114 mm & WATEC910HX & 2.0 & 1.0 & NTP & S/N low &  \makecell{ Anonymous } \\

        \hline  

        23 June, 2022 & Javalambre Observatory (Spain) & 01$^\circ$00'58.7"W & 40$^\circ$02'30.5"N & 1957.0 & 400 mm & ProLine PL4720 & 5.0 & 5.4 & NTP & Positive &  \makecell{R. Iglesias-Marzoa\\ N. Maícas \\ F. J. Galindo} \\

        \hline  
                
        23 June, 2022 & La Hita Observatory (Spain) & 03$^\circ$10'59.9"W & 39$^\circ$34'04.8"N & 770.0 & 770 mm & SBIG STL11000 & 10.0 & 12.0 & NTP & Positive &  \makecell{N. Morales \\ J. L. Ortiz \\ P. Santos-Sanz \\ F. Organero \\ L. Ana}  \\

        \hline  

        23 June, 2022 & Calar Alto Observatory (Spain) & 02$^\circ$32'44.9"W & 37$^\circ$13'24.7"N & 2173.2 & 1200 mm & DLR & 1.3 & 1.5 & GPS & Positive &  \makecell{S. Mottola}  \\ 

        \hline 
        
        23 June, 2022 & Botorrita (Spain)  & 01$^\circ$1'15.1"W & 41$^\circ$29'50.5"N & 403.0 & 500 mm & QHY174M & 3.5 & 3.5 & GPS & S/N low & \makecell{ O. Canales \\ D. Lafuente \\ S. Calavia \\ Ch. Oncins }  \\            

        \hline 

        17 May, 2023 & Shiawassee (USA)  & 84$^\circ$01'59.2"W & 43$^\circ$19'30.8"N & 178.0 & 508 mm & QHY174M & 2.0 & 2.0 & GPS & Positive &   \makecell{K. Getrost}  \\      

        \hline 
        
        17 May, 2023 & Comer Field (USA) & 105$^\circ$00'35.6 W & 40$^\circ$05'29.2"N & 1567.0 & 280 mm & QHY174 & 2.5 & 2.5 & GPS & Positive &  \makecell{V. Nikitin}   \\   

        \hline 
        
        17 May, 2023 & Nederland (USA) & 105$^\circ$26'44.0"W & 39$^\circ$59'13.9"N & 2492.6 & 200 mm & QHY174M & 2.5 & 2.5 & GPS & Positive & \makecell{ A. Verbiscer \\ M. Skrutskie }  \\    

        \hline 

        19 May, 2023 & GOA (Brazil) & 40$^\circ$19'02.0"W & 20$^\circ$18'01.9"S & 24.0 & 304 mm & ZWOASI1600MMPro & 5.0 & 6.8 & NTP & Positive &   \makecell{M. Malacarne}   \\   

        \hline 

        19 May, 2023 & Sonear3 Observatory (Brazil) & 43$^\circ$41'24.0"W & 19$^\circ$49'27.2"S & 1498.9 & 450 mm & QHY600 & 5.0 & 8.0 & NTP & Positive & \makecell{C. Jacques}    \\   

        \hline 

        19 May, 2023 & DogsHeaven Observatory (Brazil) & 47$^\circ$54'39.9"W &  15$^\circ$53'30.1"S & 1044.8 & 508 mm & ASI294MM & 5.0 & 6.8 & NTP & S/N low &    \makecell{P. Cacella}    \\      
        
       \hline 

        01 July, 2023 & Apache Point Observatory (USA) & 105$^\circ$49'13.1"W &  32$^\circ$46'48.3"S & 2788.0 & 3500 mm & Agile & 0.5 & 0.5 & GPS & Positive &   \makecell{ A. Verbiscer \\ R. DeColibus \\ A. Peck \\ C. Gray}    \\      
        
       \hline

    \end{tabular}
   \end{minipage}}
\end{table*}

\begin{table*}[t]
\centering
\renewcommand{\arraystretch}{1.5}
\caption{UT ingress and egress times and errors for each observation.}
\label{table:ing_egr_times}
\begin{tabular}{l c c}

        \textbf{Chord}      &  \textbf{Ingress time}  & \textbf{Egress time}  \\
        \hline
        \hline

        \multicolumn{1}{c}{\color{gray}Occ. A - 26 May, 2020\color{black}} \\
        Borowiec (Poland) & 22:30:07.863 $\pm$ 0.697 & 22:30:22.307 $\pm$ 0.313 \\
        Odesa-Mayaki (Ukraine) & 22:29:33.218 $\pm$ 0.736 &  22:29:50.632 $\pm$ 0.752 \\  
        Konkoly (Hungary) & 22:30:04.601 $\pm$ 0.258 & 22:30:23.218 $\pm$ 0.257\\
        Gothard (Hungary) & 22:30:16.151 $\pm$ 0.044 & 22:30:33.211 $\pm$ 0.045 \\
        Millen-Bruch (Germany)* & 22:30:42.094 $\pm$ 0.954 & 22:30:58.064 $\pm$ 1.110 \\
        Nandrin (Belgium)* & 22:30:45.845 $\pm$ 0.191 & 22:30:57.746 $\pm$ 0.185 \\

        \multicolumn{1}{c}{\color{gray}Occ. B - 23 June, 2022\color{black}} \\

        Javalambre (Spain) & 00:50:34.107 $\pm$ 1.064 & 00:50:45.536 $\pm$ 1.419 \\
        La Hita (Spain) & 00:50:36.114 $\pm$ 2.276 & 00:50:55.567 $\pm$ 2.930 \\
        Calar Alto (Spain) & 00:50:33.812 $\pm$ 0.391 & 00:50:52.265 $\pm$ 0.336 \\

       \multicolumn{1}{c}{\color{gray}Occ. C - 17 May, 2023\color{black}} \\

        Shiawassee (USA) & 08:50:42.397 $\pm$ 0.344 & 08:51:03.586 $\pm$ 0.542 \\
        Comer Field (USA) & 08:52:08.143 $\pm$ 0.652 & 08:52:24.456  $\pm$ 1.355 \\
        Nederland (USA) & 08:52:07.448 $\pm$  0.778 & 08:52:25.670 $\pm$ 0.689 \\

        \multicolumn{1}{c}{\color{gray}Occ. D - 19 May, 2023\color{black}} \\

        GOA (Brazil) & 03:10:38.132 $\pm$ 0.850 & 03:10:50.035 $\pm$ 1.345 \\
        Sonear3 (Brazil) & 03:10:52.356 $\pm$ 0.594 & 03:11:08.267 $\pm$ 0.596 \\

        \multicolumn{1}{c}{\color{gray}Occ. E - 01 July, 2023\color{black}} \\

        Apache Point (USA) & 07:09:08.865 $\pm$ 0.045 & 07:09:22.961 $\pm$ 0.047 \\

\end{tabular}

\tablefoot{*For Millen-Bruch and Nandrin, time corrections of -2.5 seconds and -0.66 seconds, respectively, were applied. These cameras exhibit certain time offsets that depend on the observation mode (a detailed analysis can be found in the following  \href{http://www.dangl.at/ausruest/vid_tim/vid_tim1.htm}{link}).}

\end{table*}

\begin{table*}
\centering
\caption{Parameters obtained through $\chi^2$ minimization for Occ. B, C, D, and E.}
\label{table:elli_parameters}
\renewcommand{\arraystretch}{1.5}
\begin{tabular}{c c c c c}

        \textbf{} & \textbf{Occ. B} & \textbf{Occ. C} & \textbf{Occ. D} & \textbf{Occ. E} \\
        
        \hline
        Center coordinate $x_{0}$ ($km$) & 23.0 & 98.3 & 50.1 & 1020.2 \\
        \hline
        Center coordinate $y_{0}$ ($km$) & 193.34 &  134.4 & 66.0 & 171.2 \\
        \hline
        Semi-major axis, $u$ ($km$) & 241.0 &  241.0 & 242.4 & 242.2\\
        \hline
        Semi-minor axis, $v$ ($km$) & 158.4 & 159.8 & 159.8 & 159.7 \\
        \hline
        Position angle, $P.A.$ ($^{\circ}$)& 107.4 & 104.8  & 104.8 & 107.4 \\
        \hline
           
\end{tabular}
\end{table*}

\end{appendix}
\end{document}